\documentclass[amssymb,showpacs,preprint,pre,aps]{revtex4}

\usepackage{amssymb}
\usepackage{graphicx}

\begin{document}

\title{Magnetostatic Spin Waves and %
Magnetic-Wave Chaos in Ferromagnetic Films. \\ %
III. Numeric Simulations of Microwave-Band Magnetic Chaos, Its Synchronization %
and Application to Secure Communication}

\author{A.A. Glushchenko, Yu.E. Kuzovlev, Yu.V. Medvedev, and N.I. Mezin}
\email{kuzovlev@fti.dn.ua}

\affiliation{A.A.Galkin Physics and Technology Institute of NASU, ul.
R.Luxemburg 72, 83114 Donetsk, Ukraine}

%\date{\today}

\begin{abstract}
Selected results of original numeric simulations of %
non-linear magnetostatic spin waves and microwave-frequency %
magnetic chaos in ferrite films are expounded, as third part of the work %
whose first two parts are recent arXive preprints 1204.0200 and 1204.2423 . %
Especially we consider crucial role of parametric processes in creating %
the chaos and simultaneously obstacles to its synchronization, %
and examine some possibilities of good enough synchronization (to an %
extent  allowing its use for direct secure communication in microwave
band).
\end{abstract}


\maketitle

%%%%%%%%%%%%%%%%%%%%%%%%%%%%%%

\section*{Introduction}

This preprint continues preprints  1204.0200 and 1204.2423 %
and represents some results of numerical simulations %
of auto-generation magnetic-wave chaos in ferrite films, its %
synchronization and application to secure communication. %
A references like (6.x) means formula (x) from Section 6 %
placed in  1204.2423\,. Sections 1-5 are placed in 1204.0200\,.  %

Our main interest below is

(i) visual investigation of %
regular and chaotic non-linear magnetostatic %
spin wave patterns auto-generated through feedback %
consisting of wire inductors (antennae), amplifier and may be %
filters;\, %

(ii) investigation of those conditions of the auto-generation, %
and properties of generated patterns, %
what are mostly responsible for characteristics of resulting %
chaotic microwave-band electric signals (voltages and currents), %
in particular, their possibilities to synchronize chaotic %
patterns in other similar systems (ferrite film) and thus  %
to serve for secure transmission of information.

\section*{7. NUMERIC SIMULATIONS OF %
AUTO-GENERATION AND SYNCHRONIZATION OF %
MAGNETIC-WAVE CHAOS}

7.1. CHAOTIC AUTO-GENERATION IN FILM WITH LINEAR FEEDBACK.

If the voltage (EMF) signal, \,$\,U(t)\,$\, , induced in a conductor
(antenna) by
magnetization precession is amplified and transformed into driving current, \,$\,%
J(t)\,\,$\,\ , in another conductor, then auto-generation of MW may
take place. The simplest variant of this feedback is drawn at plot
(A) in Fig.13. It consists of two identical loop inductors and purely
linear amplifier whose gain and phase shift are with frequency
independent. Typical features of chaotic generation in such the
scheme, at \,$\,H_{0}=2\,\ \,$\,, are presented by Figs.13 and 14.

The plots (B) and (E) in Fig.13 show that rather wide-band
magnetization chaos is produced, in frequency range about 600 MHz.
Position of most intensive peak at the voltage spectrum well
corresponds to frequency of the Damon-Eshbach (DE) surface wave with
length \,$\,\ \lambda =2l=\,$\, \,$\,32D\,$\, dictated by the loop
width, i.e. to \,$\,\,\omega _{DE}=\,$\, \,$\,\ \omega _{DE}(2\pi
D/\lambda )\,\,$\,\ (\,$\,\approx 2.54\,\,$\,\ GHz at\,$\,\,\,$\,
\,$\,H_{0}=2\,\,$\,). The lower edge of the
spectrum lies below uniform precession frequency\thinspace \,$\,\,\omega _{u}\,\,$\,%
\ (\,$\,\approx 2.12\,\,$\,\ GHz \,$\,\,\,$\,). This means that the
main-branch bulk waves also are excited. The upper edge\ of the
voltage spectrum is far below the upper frequency of MSW,
\,$\,(H_{0}+2\pi )f_{0}\,\,\,$\,\ (\,$\,\approx \,$\,
\,$\,3.25\,\,$\,\ GHz). Hence, long surface MSW are dominating in
magnetization patterns.

This pattern is seen at plot (A) in Fig.14. The contour plot of its
spatial spectrum ((F) in Fig.14) demonstrates tracks of (i) a set of
long-wave modes with frequencies about \thinspace \,$\,\,\omega
_{u}\,\,$\,\ , \ (ii) three times shorter than \,$\,\,\lambda \,$\,
DE mode with frequency \,$\,\ \omega _{DE}(3\cdot 2\pi D/\lambda
)\,\,$\,\ (\,$\,\approx 2.95\,\,$\,\ GHz ), \ (iii) comparatively
short-wave surface modes with non-zero transversal wavenumber,
\,$\,k_{y}\neq 0\,$\, , and frequencies about\thinspace\ \,$\,\omega
_{DE}(2\pi D/\lambda )\,$\, , and, besides, (iv) approximately two
times longer than \,$\,\,\lambda \,$\, DE mode with frequency
\,$\,\,\approx \omega _{DE}(\pi D/\lambda )\,\,$\,\ ( \,$\,\approx
2.36\,\,$\,\ GHz).

\begin{figure}
\includegraphics{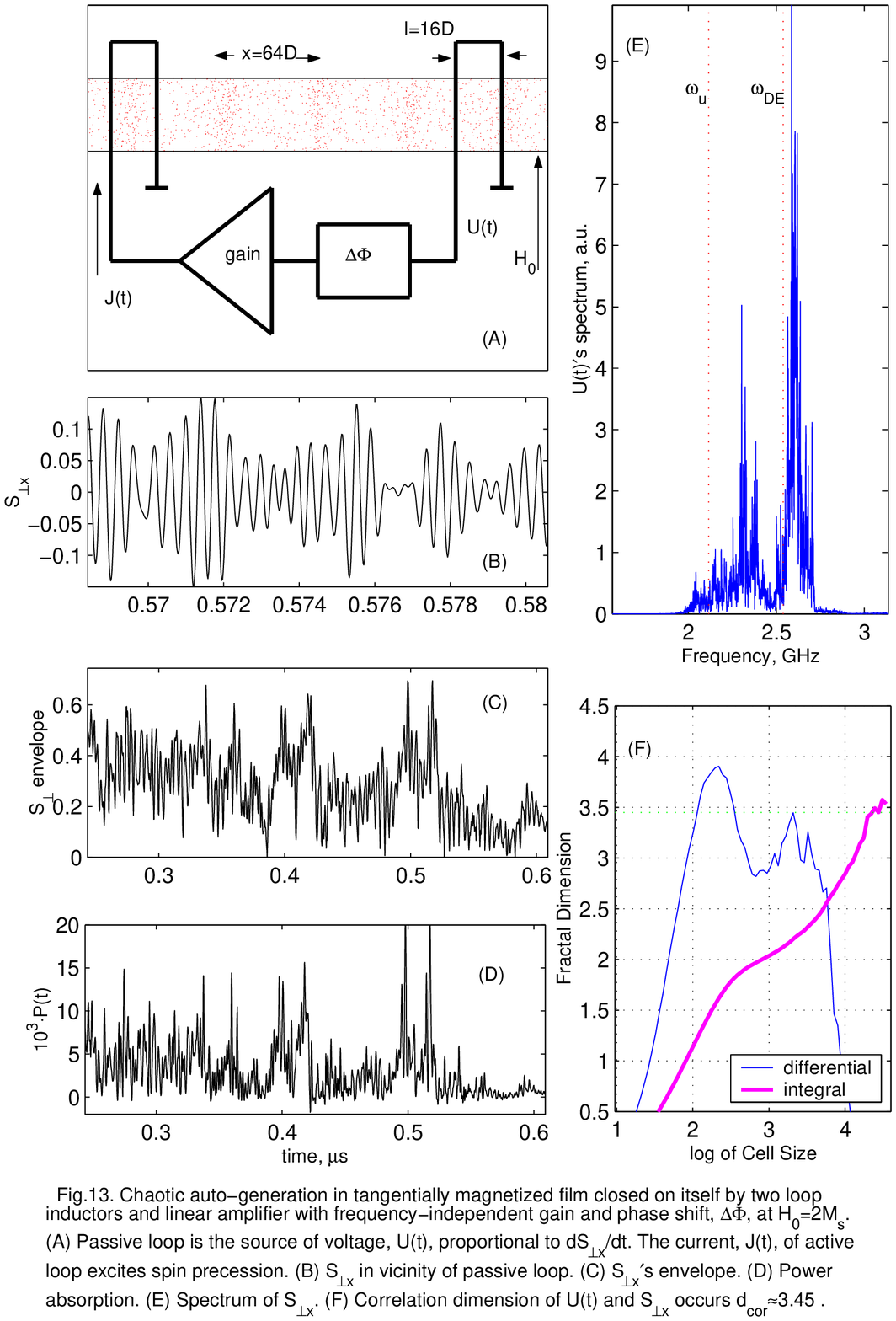}
\end{figure}

\begin{figure}
\includegraphics{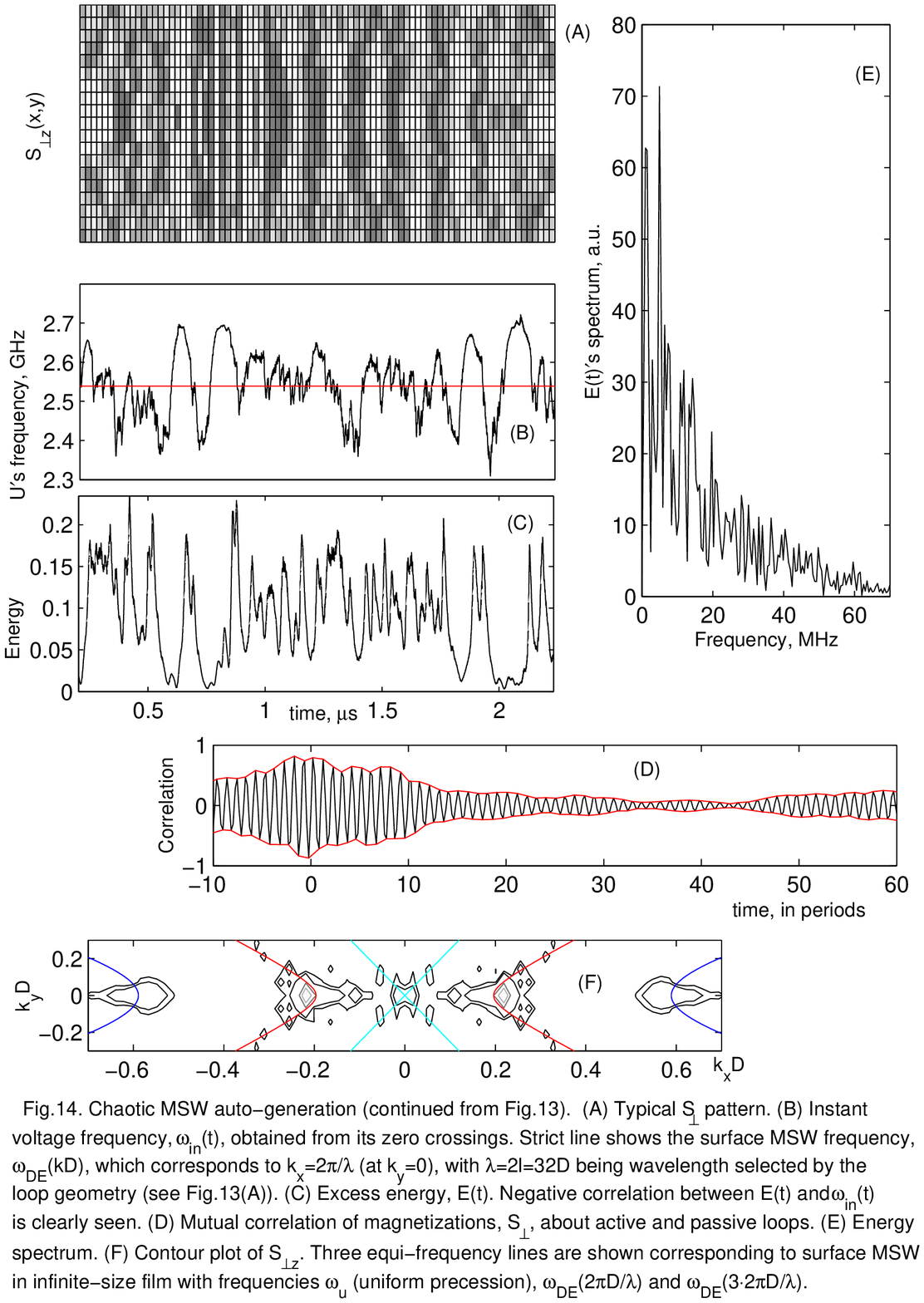}
\end{figure}

Interestingly, the latter mode seems be responsible for lower high peak in
the voltage spectrum. Plots (B) and (C) show that, in rough terms,
auto-generation switches between two states whose frequencies are about \,$\,%
\omega _{DE}(2\pi D/\lambda )\,$\, and \,$\,\omega _{DE}(\pi
D/\lambda )\,$\, , and that at the first of them the energy is
usually lower than at the second (in accordance with negative
non-isochronity, see Sec.5).

Due to linearity of amplifier, the film itself are forced to take all
cares about nonlinear saturation of precession angle. Naturally, this
results in complicated large-amplitude chaos characterized by wide
variety of time scales and intermittency of different regimes, which
is illustrated by plots (B)-(D) in Fig.13 and (B),(C) and (E) in
Fig.14. Low-frequency contents of this chaos is better visible in the
spectrum of energy, \,$\,E(t)\,\,$\,, which is enriching at least
down to 5 MHz. Voltage and magnetization time series yield fractal
dimension \,$\,d_{cor}\approx 3.45\,$\, . Hence, we encounter
hyper-chaos governed by not less than four relevant variables.

7.2. AUTO-GENERATOR WITH NON-LINEAR FEEDBACK.

As practice says, in real generators an inner non-linearity of
amplifiers is essential. It manifests itself in two ways: (i) as
non-linearity of input resistance which effectively saturates input
voltage, and (ii) as saturation of output power or, equivalently, of
output driving current. Besides, one should take into account
frequency dependence of gain, \,$\,K(\omega )\,\,$\,, and own
impedances (capacities and inductances) of feedback conductors.

Corresponding scheme looks as in Fig.15. Here \,$\,R_{0}\,\,$\,, \,$\,L_{0}\,\,$\,and \,$\,%
C_{0}\,\,$\,\ are linear (small-amplitude) input resistance,
\,$\,\sim 50\,\,$\,\ Ohm, inductance of passive (feedback) antenna,
\,$\,\sim 5\,\,$\,\ nH, and capacity of active (driving) antenna,
\,$\,\sim 0.5\,\,$\,\ pF, respectively. Since in reality such the
circuit is quasi-stationary one (its size is much smaller than
lengths of EM-waves under operation), a simple analysis shows that
there is no necessity in additional reactive elements. For practical
high-frequency wide-band amplifiers, typical level of the input
saturation is \,$\,U_{sat}\sim \,$\, \,$\,1\div 3\,\,$\,\ V\thinspace
,\ level of output saturation is \,$\,J_{sat}\,$\, \,$\,\sim 20\div
70\,\,$\,\ mA\thinspace , while \,$\,\,K_{0}=\,$\, \,$\,\max
\,|K(\omega )|\,$\, \,$\,\sim 5\div 15\,\,$\,, and linear frequency
characteristics can be modeled by

\begin{equation}
K(\omega )=K_{0}\omega _{0}^{2}(\omega _{0}^{2}-\omega ^{2}-ig\omega
)^{-1}\exp (i\omega \tau _{d})\,\ ,
\end{equation}
with \,$\,\tau _{d}\,$\, describing time delay. In numerical
simulations, usually
values\thinspace\ \,$\,\omega _{0}/2\pi =3.4\,\,$\,\ GHz \ and \,$\,g=6\cdot 10^{9}\ \,$\,s\,$\,%
^{-1}\,\,$\,were used, as in INA-32063 wideband silicon RFIC
amplifier ( \,$\,\tau _{d}\,$\, is small as compared with time delays
in film and therefore insignificant).

\begin{figure}
\includegraphics{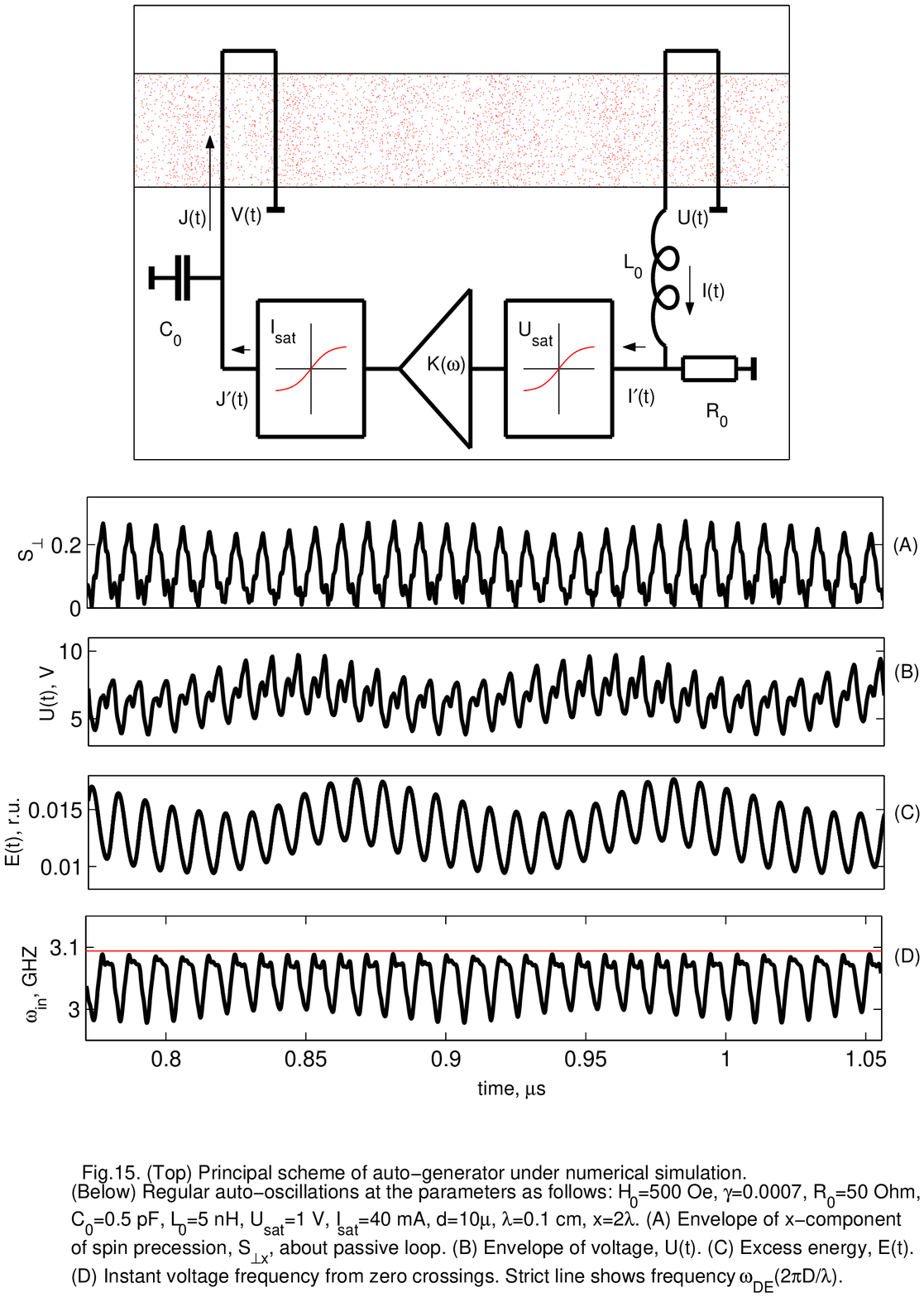}
\end{figure}

Fig.15 shows how numerical copy of this scheme generates at
relatively small gain, \,$\,K_{0}\approx 5\,\,$\,, and moderate
output power saturation. We see that likely periodic (although more
or less complicated) oscillations take place in all the amplitudes
and phases. If decreasing gain and /or output saturation level one
may observe more simple oscillations, down to trivial mono-chromatic
regime and failure of generation, while increase results in further
complications and then chaos.

7.3. ROLE OF ANTENNAE SEPARATION.

The Fig.16 illustrates chaotic auto-generation in numerical model
corresponding to \,$\,4.4\,\,$\,mm\thinspace \,$\,\times
0.9\,\,$\,mm\thinspace \,$\,\times 10\,\mu \,$\,m\thinspace -size
film and \,$\,\,l=0.5\,\,$\,mm width of loop inductors,
or to a proportionally re-scaled system (but not too small), at field \,$\,%
H_{0}=500\,\,$\,Oe , moderate gain and relatively high output power
saturation. At top of Fig.16 static magnetization of such shaped
film, at \,$\,M_{s}=140\,\,$\,\ Oe\ and \,$\,H_{0}=500\,\,$\,Oe , is
presented. Clearly, a noticeable demagnetization takes place at film
corners only.

In this numerical experiment, two variants of separation, \,$\,x\,$\,
, of feedback
and active inductors were considered, with \,$\,\,\,x=\,$\, \,$\,x_{1}\,$\, \,$\,=2\lambda \,\ \,$\,%
and \,$\,\,x=\,$\, \,$\,x_{2}=\,$\, \,$\,1.8\lambda \,\,$\,\ , where \,$\,\,\lambda =2l=1\,\,$\,%
mm\thinspace\ is maximum of wavelengths preferably selected by the
loop antennae. Comparison of these cases demonstrates, at plot (D),
that auto-generation spectrum is sensitive not only to the
\,$\,\,\lambda \,\,$\,\ but also to \,$\,\,x\,\,$\,. It could be
expected, because any phase shift in feedback should be compensated
by equal magnetic wave phase difference, \,$\,\,k_{x}x\,$\, , between
antennae, with \,$\,\,k\,$\, being a dominating wave vector. At
smooth feedback frequency characteristics, one can suppose that
\,$\,k_{1x}x_{1}\approx k_{2x}x_{2}\,\ \,$\,\ , therefore

\begin{equation}
\,\,k_{2x}-k_{1x}\approx -k_{1x}(x_{2}-x_{1})/x_{1}\,,\,\ \ \,\omega
_{2}-\omega _{1}\approx v_{g}(k_{2x}-k_{1x})\,\,,
\end{equation}
where \,$\,\,\omega _{2}-\omega _{1}\,$\, is corresponding change in frequency and \,$\,%
v_{g}\,\,$\,\ is group velocity. According to this estimate, transition from \,$\,%
\,\,x_{1}=2\lambda \,\ \,$\,to \,$\,\,x_{2}=1.8\lambda \,\,$\,\ must
result in increase of auto-generation frequencies, which agrees with
numerical simulation.
Taking \,$\,\,\omega \,\,$\,\ be the frequency of DE wave, we obtain\thinspace\ \,$\,%
v_{g}\approx \,$\, \,$\,4\pi ^{2}D/\omega _{1}\,$\, and
then\thinspace\ \,$\,\omega _{2}-\omega _{1}\approx \,$\,
\,$\,0.032\,\,\,$\,\ (in dimensionless units), that is\thinspace\
\,$\,\approx 13\,\,$\,\ MHz , in satisfactory quantitative agreement
with plot (D).

Let us more discuss the voltage spectrum. As in case of linear feedback (see
above), its upper-frequency peak can be surely related to DE mode with \,$\,%
k_{x}\approx 2\pi /\lambda \,\,$\,. But what is origin of the
lower-frequency peak? From watching for dynamics of magnetization
pattern, it is possible to
relate this peak to a long DE mode with wavelength \,$\,\,nx\,\,\,$\,\ where \,$\,%
1\lesssim n\lesssim 2\,\,$\,. Such mode also can get suitable phase
conditions under feedback determined by Eq.1.

\begin{figure}
\includegraphics{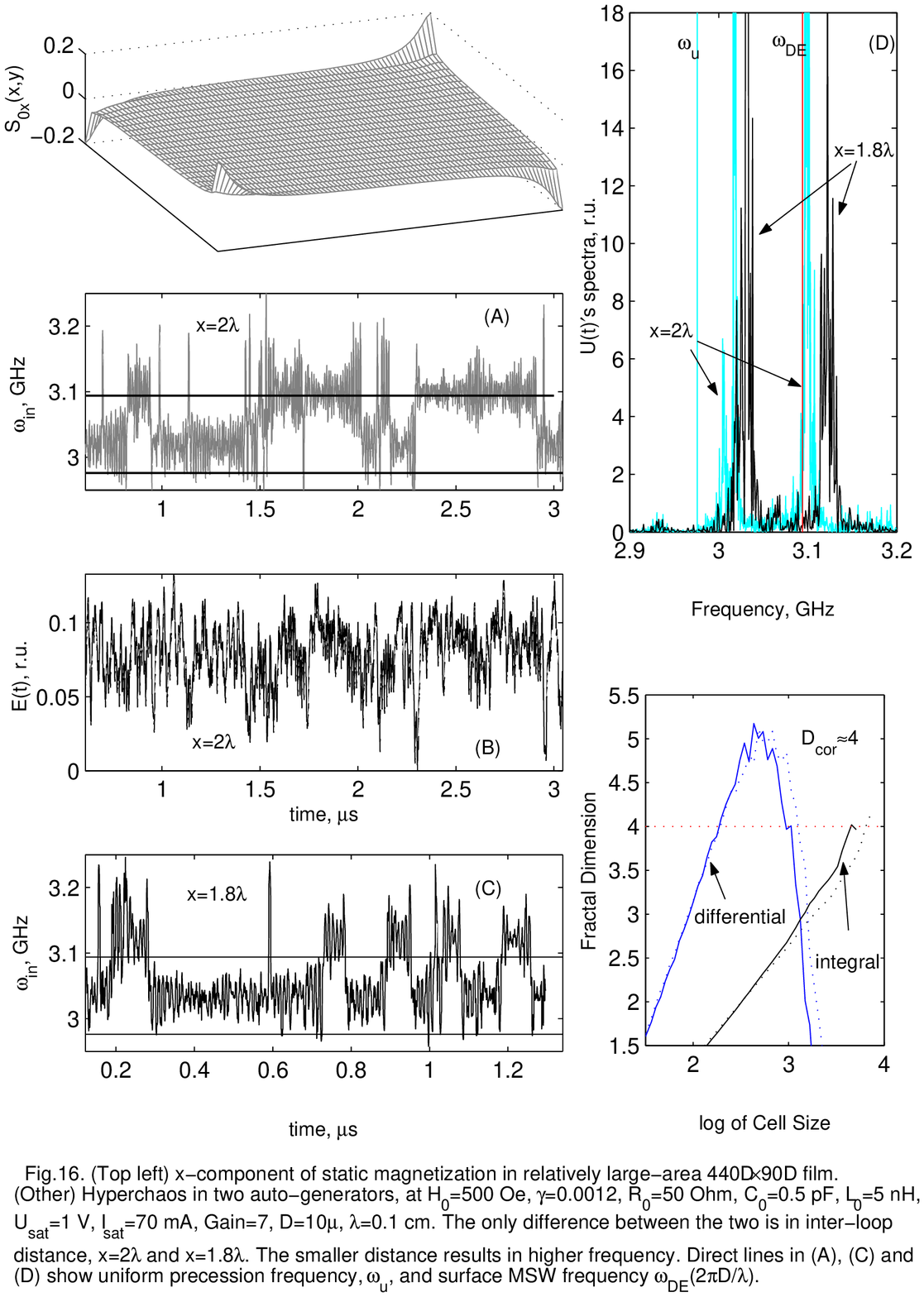}
\end{figure}

7.4. HYPER-CHAOS UNDER NON-LINEAR FEEDBACK.

Non-linear concurrence between the two dominating wave modes
manifests itself in chaotic jumps of instant (time-local) frequency
of auto-generation, \,$\,\,\omega _{in}(t)\,$\, , as shown by plots
(A) and (C) in Fig.16. It should be underlined that from the point of
view of energy pattern (quasi-local energy, see Sec.6.14) the same
interplay of different modes looks as the birth and drift of envelope
solitons. In these terms, the frequency separation of two main peaks
at plot (D) is nothing but mean number of soliton births per second
(see below).

Due to high power saturation level, the whole picture occurs very
chaotic. Its correlation dimension obtained from voltage and
magnetization time series (solid and dot lines, respectively) is
\,$\,d_{cor}\approx 4\,\,$\,\ (if not greater).

7.5. WEAK CHAOS IN\ GENERATOR WITH WIRE INDUCTORS.

It is interesting how the picture will be changed if two-element loop
inductors are replaced by wires, at the same feedback circuit.
Figs.17a and 17b can give the answer for case
\,$\,x=150D\,$\,\thinspace\ (=1.5 mm) with \,$\,\,x\,\,$\,\ being
distance between the wires.

In this case, two dominating modes chosen by the system are long DE
waves with lengths\thinspace\ \,$\,\approx 2x\,\,\,$\,\ and
\,$\,\,\approx x\,\,$\,. In spatial
spectrum of magnetization (in momentum space) the mode with length \,$\,%
\,\approx x/2\,\,$\,\ can be detected, but practically it does not
contribute to voltage spectrum. Besides, as usually, a set of
comparatively weak excited equal-frequency modes with \,$\,k_{y}\neq
0\,\,$\,\ is present. In real space non-linear magnetization pattern
characteristic rhombic structuring is well visible invoked by
dispersing properties of underlying linear MW eigenmodes. But
frequencies of the latter become lowered by non-linearity.

At nearly the same output saturation and even greater linear gain,
the whole picture is significantly more smooth and less chaotic than
in case of loop inductors, with \,$\,d_{cor}\approx 2.55\,\,$\, only!
In part, the reason is that both the EMF accepted by wire inductor
and power pumped by it are approximately two times smaller than by
loop inductor. In this concrete experiment, power absorption
\,$\,\,P\approx \,$\, \,$\,\left\langle JV\right\rangle _{T}\sim
\,$\, \,$\,25\,\,$\,\ mW\thinspace\ , where \,$\,J\,\,$\,\ and
\,$\,V\,\,\,$\,\ are driving current and EMF in active inductor,
respectively, and \,$\,\left\langle ..\right\rangle _{T}\,\,$\,\
means time average (over a few precession periods).

7.6. GENERATION\ OF ENVELOPE SOLITONS.

In the middle plot in Fig.17b, typical spatial distribution of the
quasi-local energy density (Sec.6.14) is shown (to be precise, of \,$\,%
\,e_{loc}/H_{0}\,$\, where \,$\,H_{0}\,\,$\,\ is
\,$\,\,M_{s}\,\,$\,units). It highlights presence of three envelope
solitons of whose the first (most left) is in birth, the second is
just passing through feedback inductor and the third is dying
remainder of previously passed soliton. Obvious rhombic shaping of
the solitons reflects four most preferred directions of local energy
propagation
(corresponding to maximum group velocity), \,$\,\,|v_{gy}|/\,$\, \,$\,|v_{gx}|\approx \,$\, \,$\,%
\sqrt{H_{0}/4\pi }\,\,$\,\ (see Sec.5.4).

\begin{figure}
\includegraphics{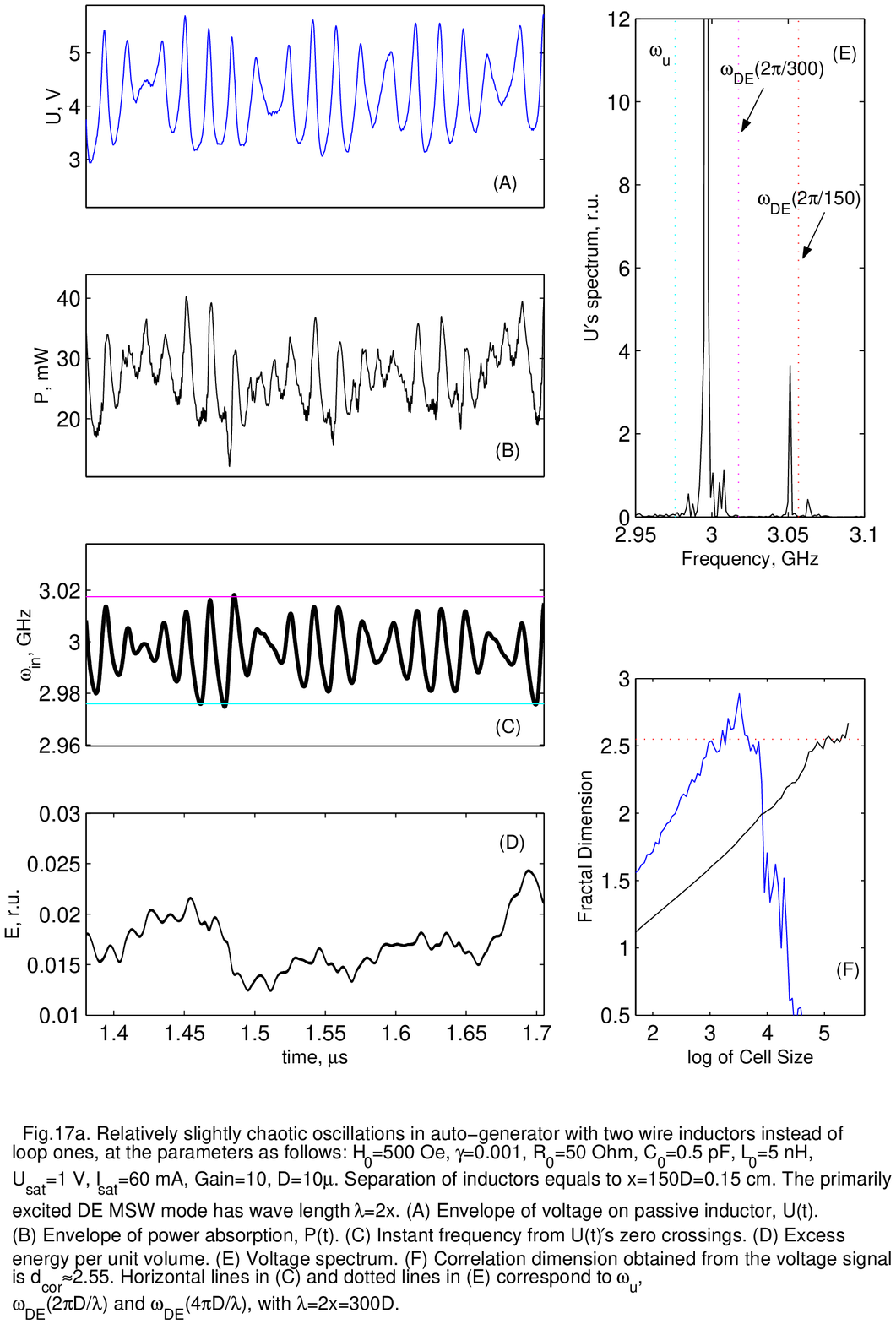}
\end{figure}

\begin{figure}
\includegraphics{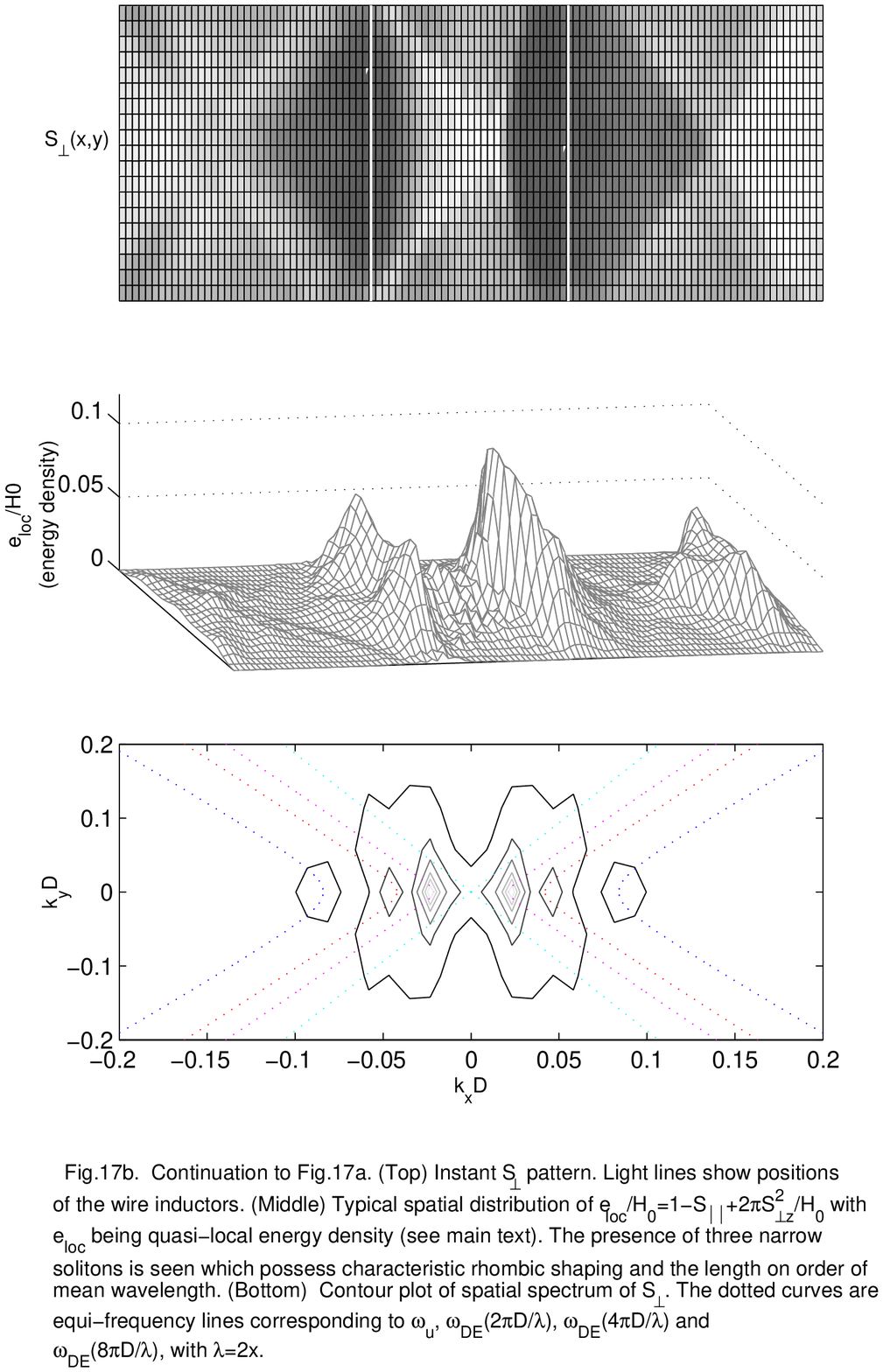}
\end{figure}

Notice that solitons are very sharp (as measured in
\,$\,x\,$\,-direction): their width is even less than wavelengths of
carry MSW enveloped by them! But several wavelengths have time to
pass through such the envelope while it itself displaces on its
width.

7.7. INSTANTON STRUCTURING OF SPIN PRECESSION.

Let us go back to strong chaos considered above in subsection 4 and
have a look how envelope solitons manifest themselves in local
magnetization time series. The Fig.18 shows behavior of spin and of
\,$\,\,e_{loc}/H_{0}\,\,$\,\ , where \,$\,\,e_{loc}\,$\, is
quasi-local energy density (Sec.6.14), at a point close to the
feedback inductor. Notice that

\begin{equation}
\,e_{loc}/H_{0}\approx (S_{x}^{2}+p^{2}S_{z}^{2})/2\approx p|\Psi |^{2}/2\,\
,
\end{equation}
where \,$\,\,\Psi \,\,$\,\ is the wave function (see in Sec.5.3 and
Eq.5.9 there), and \,$\,\,p\,\,$\,\ is the eccentricity (see Eqs.4.6
and 5.10), i.e. can be interpreted as merely squared angle of
precession. Comparison of plots (C) and (D) in Fig.18 visually prove
that at fast time scale (measured in periods of precession)
\,$\,\,e_{loc}\,\,$\,is almost integral of motion.

What most of all impresses in plots (B), (C) and (F) is that \ \,$\,%
\,\,e_{loc}\, \,$\, (and thus local angle of precession) never
overlooks to
regularly turn into either exact zero or to extremely small value. If \,$\,%
\,S_{\bot }(r,t)\,\,$\,\ was mere random field (like thermally
activated magnetization noise in thermodynamical equilibrium) such
behavior would be absolutely improbable. It gives best evidence that
we observe results of essentially nonlinear self-organization of
magnetization field in ``clots''. As in nonlinear field theory,
spatially local time tracks of self-organization may be termed
instantons.

As it was mentioned in Sec.5.4, to be more precise, the eccentricity, \,$\,\,p\,\,$\,%
, should \ be related to dominating MSW mode (instead of uniform one). In
other words, the quasi-local energy introduced in Eq.6.6 must be redefined
with taking into account non-local (gradient dependent) part of energy
estimated by Eq.6.7. Corresponding refinement reads

\begin{equation}
e_{loc}=H_{0}(1-S_{y})+\pi Dk_{0}S_{x}^{2}+\pi (2-Dk_{0})S_{z}^{2}\propto
1-S_{y}+H_{m}S_{z}^{2}/2H_{0}\approx (S_{x}^{2}+p^{2}S_{z}^{2})/2\ \ \,,\,\
\
\end{equation}

\[
\frac{H_{m}}{H_{0}}\equiv \frac{4\pi (1-Dk_{0})}{H_{0}+2\pi Dk_{0}}\,\ \
,\,\ \ \ \ \ \ p=\left[ 1+\frac{H_{m}}{H_{0}}\right] ^{1/2}
\]
where \ \,$\,k_{0}\;\,$\,is module of dominating wave-vector (for loop inductors, \ \,$\,%
|k_{0}|\approx \pi /l\,\,$\,\ ) and the condition \,$\,\,Dk_{0}\ll
1\,\,$\,\ is taken in mind. For comparatively short-wave chaos, this
is better invariant of fast motion (precession) than what defined by
(6.6).

\begin{figure}
\includegraphics{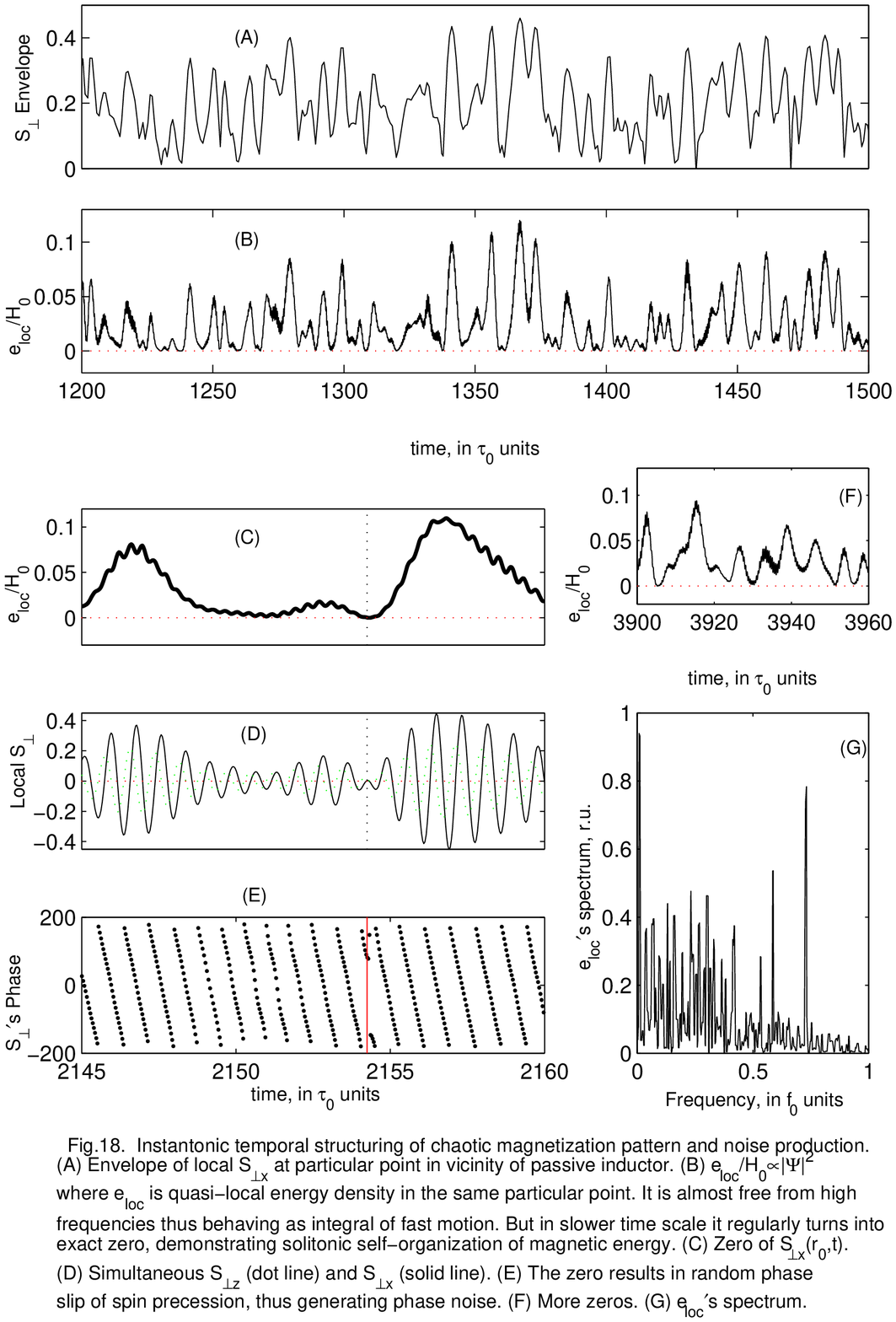}
\end{figure}

\begin{verbatim}
\end{verbatim}

7.8. PHASE SLIPS AND\ NOISE PRODUCTION.

Plots (D) and (E) in Fig.18 show that when local amplitude of precession is
passing through zero its phase can slip by some angle. In contrast to ideal
black solitons (see Sec.5.6), this angle can take any value different from \,$\,%
\pm \pi \,$\, . Of course, usually two close spins undergo the same
slip, but sometimes one or another spin can be beaten from common
behavior. Hardly this fine detail of magnetization dynamics is
unambiguously determined by a few relevant chaotic variables.
Therefore, local phase slips may become the source of some amount of
noise, in addition to chaos.

Although, by our observations, this noise is rather weak, probably it may
create obstacles to ideal copying of chaos under synchronization.

7.9. QUASI-PERIODIC CHAOS.

Fig.19 illustrates example of quasi-periodic chaos obtained in
auto-generator with relatively narrow loop antennae
(\,$\,\,l=0.2\,\,$\,\ mm) at moderate gain (=9) and output saturation
(\,$\,\,I_{sat}=55\,\,$\,\ mA), and antennae separation
\,$\,x=4l\,\,$\,. Dominating wave-length of magnetization pattern is
close to \,$\,\,\lambda =2l\,\,$\, while carrier frequency of its
oscillations is rather close to frequency of DE wave with\thinspace\ \,$\,%
k_{x}=2\pi /\lambda \,\,$\,\ and \,$\,\,k_{y}=0\,\,$\,.

In this case, time series of different variables (magnetic energy,
magnetization and voltage envelopes and phases, instant frequency,
etc.) possess characteristic features of the kind of chaotic
attractor termed noisy limit circle. The voltage spectrum at plot (F)
resembles ruled spectra of complex periodic signals (with period
being inverse distance between neighboring spectral lines).
Nevertheless, in fact voltage signal is chaotic and its fractal
dimension, \,$\,\,d_{cor}\approx 3.25\,\,$\,\ (see plot(G)) even
indicates hyperchaos.

Plot (D) shows that in spite of large magnitude of instant frequency
oscillations (\,$\,\sim 500\,\,$\,\ MHz) and therefore very wide
total frequency band of generated signal (\,$\,\approx 0.9\,\,$\,\
GHz) its phase modulation lies between well certain boundaries not
exceeding \,$\,\approx 100^{o}\,\,$\,. From comparison of plots (D)
and (E) it is seen that positive (negative) slope of time-smoothed
phase trajectory corresponds to smaller (greater) energy, in
accordance with negative non-isochronity of MW.

\begin{figure}
\includegraphics{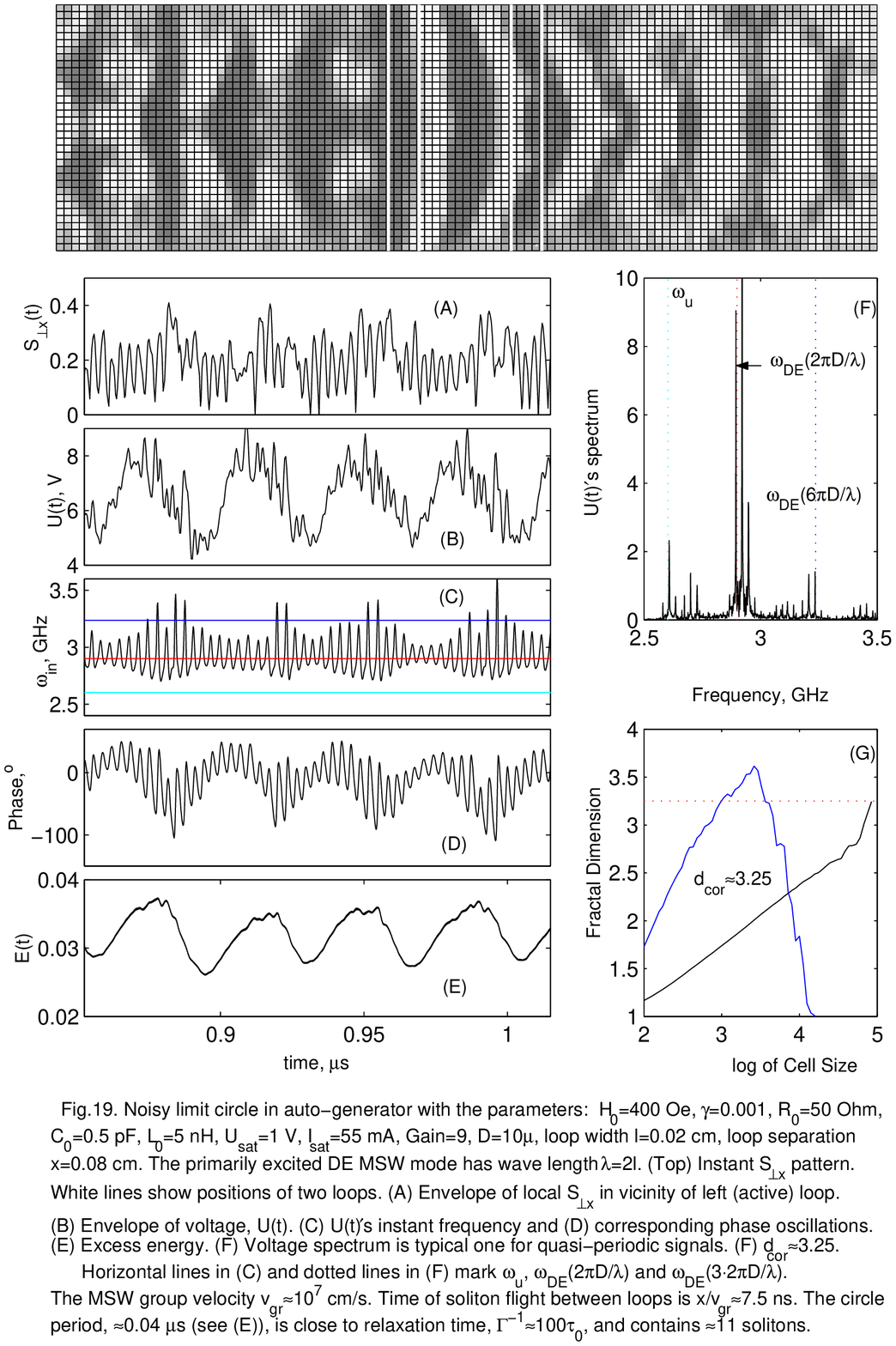}
\end{figure}

What here happens? The frequency separation of the three main groups
of lines in plot (F), \,$\,\approx 270\,\,$\,MHz, is by its sense the
frequency of soliton births near active inductor. Hence, one envelope
soliton is born approximately each \,$\,\,3.7\,\,\,$\,ns. The group
velocity of DE waves at operational frequency (see Eq.5.14) is
\,$\,\,v_{g}\approx 10^{7}\,\,$\,\ cm/s,
therefore time of soliton flight from active to passive loop is about \,$\,\,8\,\,$\,%
\ ns. Mean duration of one cycle of the quasi-periodic modulation can
be estimated from plot (E) as \,$\,\,\approx \,40\,\,$\,\ ns.
Comparing these numbers we conclude that two solitons simultaneously
are in action and that each cycle contains birth and transfer of
\,$\,10\div 12\,\,$\,\ solitons. What is for the cycle duration
itself, seemingly it can be connected to the relaxation time,
\,$\,\,\Gamma ^{-1}\,$\,.

7.10. ROLE OF\ FILM WIDTH TO WAVE LENGTH RATIO.

Intuitively, one may predict that formation and propagation of
envelope solitons is sensitive to the ratio \,$\,\,w/\lambda
\,\,$\,where \,$\,\,w\,\ \,$\,is width of film and \,$\,\,\lambda
\,\,$\,\ ( \,$\,\,\lambda \approx 2l\,\,$\,\ for two-element loop
antennae) \,$\,\,\,$\,is dominating wave length in magnetization
pattern. The Fig.20a shows that, indeed, this is confirmed by
numerical simulation of auto-generation in narrow (in the sense that
\,$\,\,w\lesssim \lambda \,\,$\,) and wide ( \,$\,\,w>\lambda
\,\,$\,) films.

In narrow film, soliton takes all width of a film. In wide film, at very
beginning of next soliton manufacturing one energy clot may be inflated
occupying all the width but soon it breaks into two or three (or even more)
solitons each of which takes a part of the width only. Then solitons run not
in parallel to film boundaries but along one of two easy propagation
directions which correspond to maximum group velocity (see Sec.5). Fig.20a
demonstrates this for the case when two soliton chains are created (of
course, if three or five soliton chains realize then central of them moves
in parallel to boundaries).

Importantly, in general in wide film (at \,$\,\,\,w/\lambda \,\gtrsim 3\,$\,%
)\thinspace\ wave fronts (equal-phase lines) of magnetization pattern
stronger are curved then in narrow film (especially in region where
different soliton chains meet one another). Therefore, the wider is film the
smaller is voltage (EMF) per unit length induced in direct antenna, up to
that wider film can produce lower voltage (at the same precession angle)
which worsens the feedback.

As example of unpretentious situation, Fig.21 presents movie of one cycle of
chaotic soliton generation in moderately pumped narrow film.

7.11. ABOUT COHERENCE OF AUTO-GENERATION.

It is known that spectrum of any oscillation with arbitrary amplitude
modulation and arbitrary (let random) but limited phase modulation consists
of infinitely narrow line (at carrier frequency) surrounded by some
pedestal. An unlimited phase modulation that is accumulation of phase slips
only produces a broadening of the carrier line.

\begin{figure}
\includegraphics{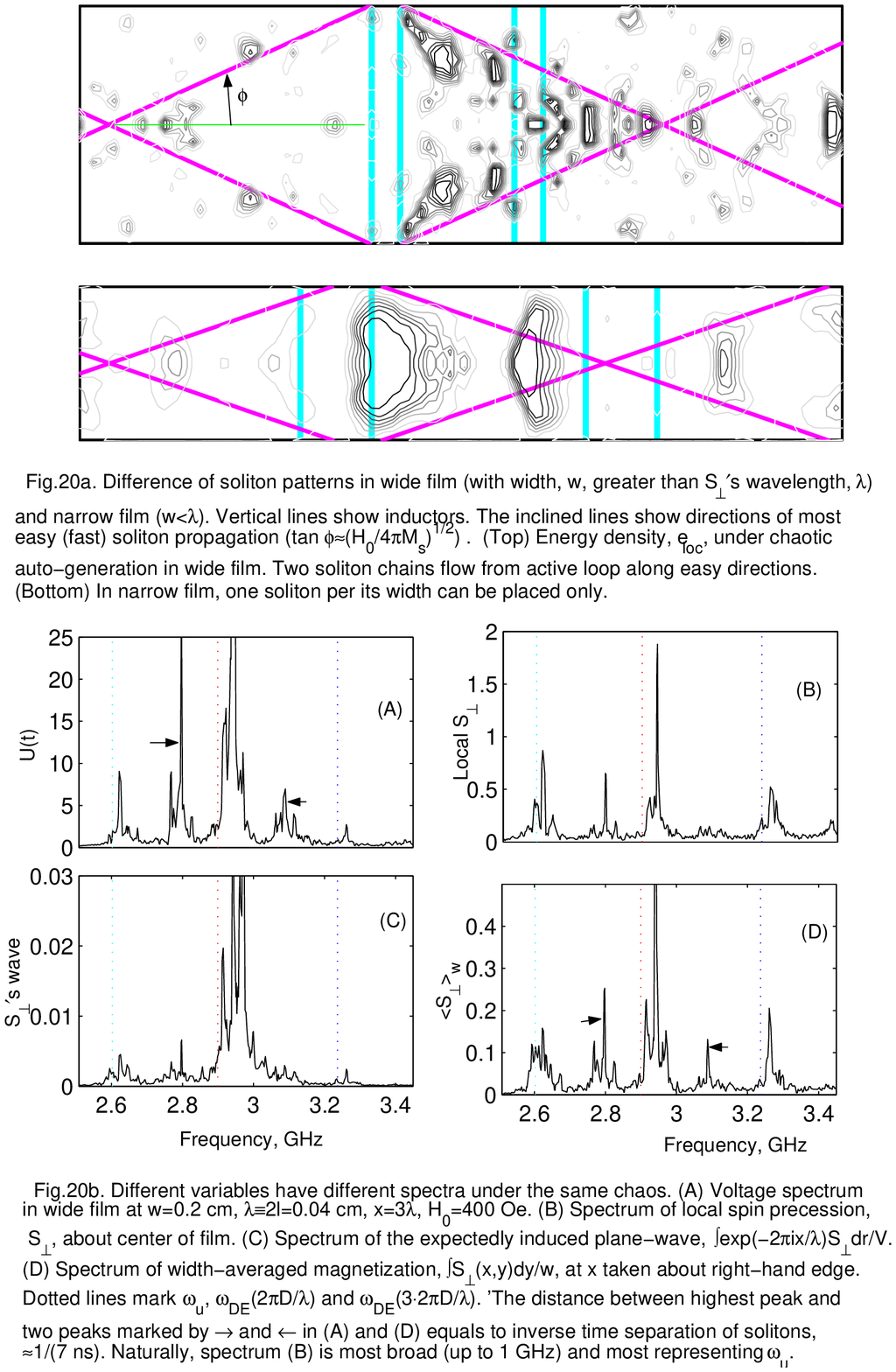}
\end{figure}

In our numerical simulations of the auto-generation (even concerning
strongly chaotic regimes) accumulation of phase slips (phase diffusion) on
the average did not exceed one-two periods of carrier frequency per
microsecond. Hence, coherent generation was observed, to the extent of
calculation duration (apparent broadening of carrier line on spectrum plots
is artifact of short-time Fourier transform).

Theoretically, the degree of coherence of chaotic generation is interesting
open question. From practical point of view, it may be better answered in
real experiments which reflect realities (like temperature instability) not
accounted for by numerical model.

7.12. SPATIAL\ DIFFERENTIATION\ OF\ AUTO-GENERATION\ SPECTRUM.

Under auto-generation, magnetization in the whole film oscillates
coherently. Due to chaotic soliton production, local phases of common
oscillation undergo chaotic deviations which however remain limited
(usually not greater then \,$\,\,\pm 60^{o}\,\,$\,). Local phase
slips do not destroy common coherent picture.

But, of course, details of pedestal of oscillation spectrum may strongly
depend on what variable is under measuring. This statement is illustrated by
Fig.20a.

The Fig.20a corresponds to wide film, with \,$\,\,\,w/\lambda =5\,$\,
and comparatively large separation of loop inductors,
\,$\,\,x=3\lambda \,\,$\,. It presents example of rather wideband
chaos with complicated spectrum which in 6 peaks can be resolved.
Again, the distance between central (most powerful) carrier peak and
its closest neighbors equals to inverse mean time separation of
soliton births. Approximately twice larger distance between the
carrier and next two neighbors can be interpreted as second harmonics
of the births frequency. Naturally, the carrier peak is most brightly
represented in spectrum of spatial Fourier transform of magnetization at \,$\,%
\,k=\{2\pi /\lambda ,0\}\,\,$\,\thinspace (plot (C)) and least
brightly in spectrum of local spin precession (plot (B)).

7.13. CIRCUMFERENTIAL\ SYNCHRONIZATION OF AUTO-GENERATION.

Naturally, the question arises about possibility of synchronization
of chaotic auto-oscillations in one generator by voltage signal taken
from another (at least identical) generator. Firstly, we tried to use
voltage (EMF), \,$\,u_{M}(t)\,\,$\,, of additional wire inductor
placed at right-hand end of the master film (i.e. behind the feedback
inductor). This voltage was compared with voltage,
\,$\,u_{S}(t)\,\,$\,, produced in identical control inductor in the
slave film. The difference, \,$\,u_{S}-u_{M}\,\,$\,, was transformed
into current, \,$\,\,J_{syn}(t)\,\,$\,, passed through one more wire
inductor close to the control one, as shown in Fig.22. Notice that in
case of exact
coincidence, \,$\,u_{S}-u_{M}=0\,$\, , the synchronizing current turns into zero, \,$\,%
J_{syn}=0\,\,$\,, and thus slave film feels no difference from master
film.

\begin{figure}
\includegraphics{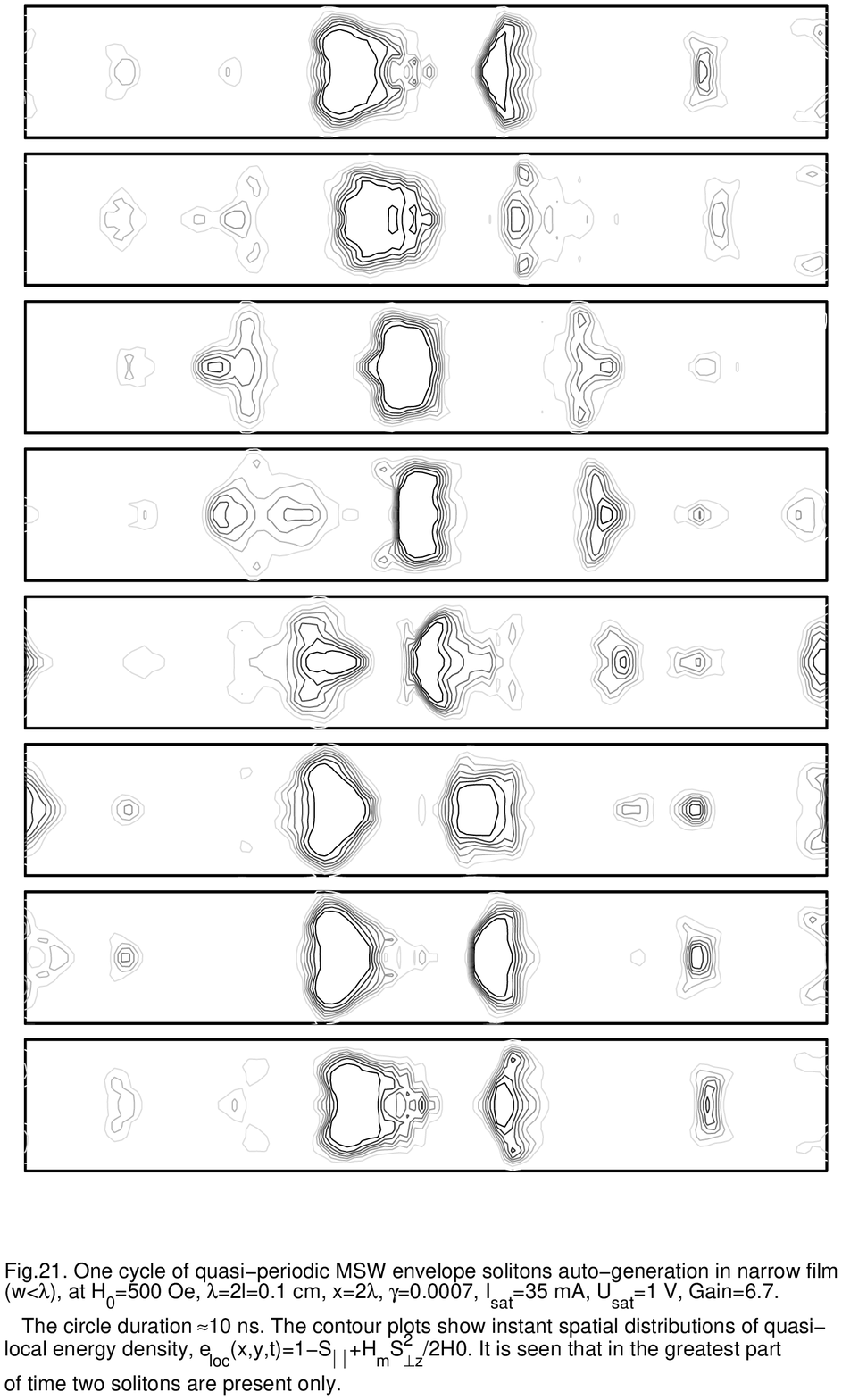}
\end{figure}

\begin{figure}
\includegraphics{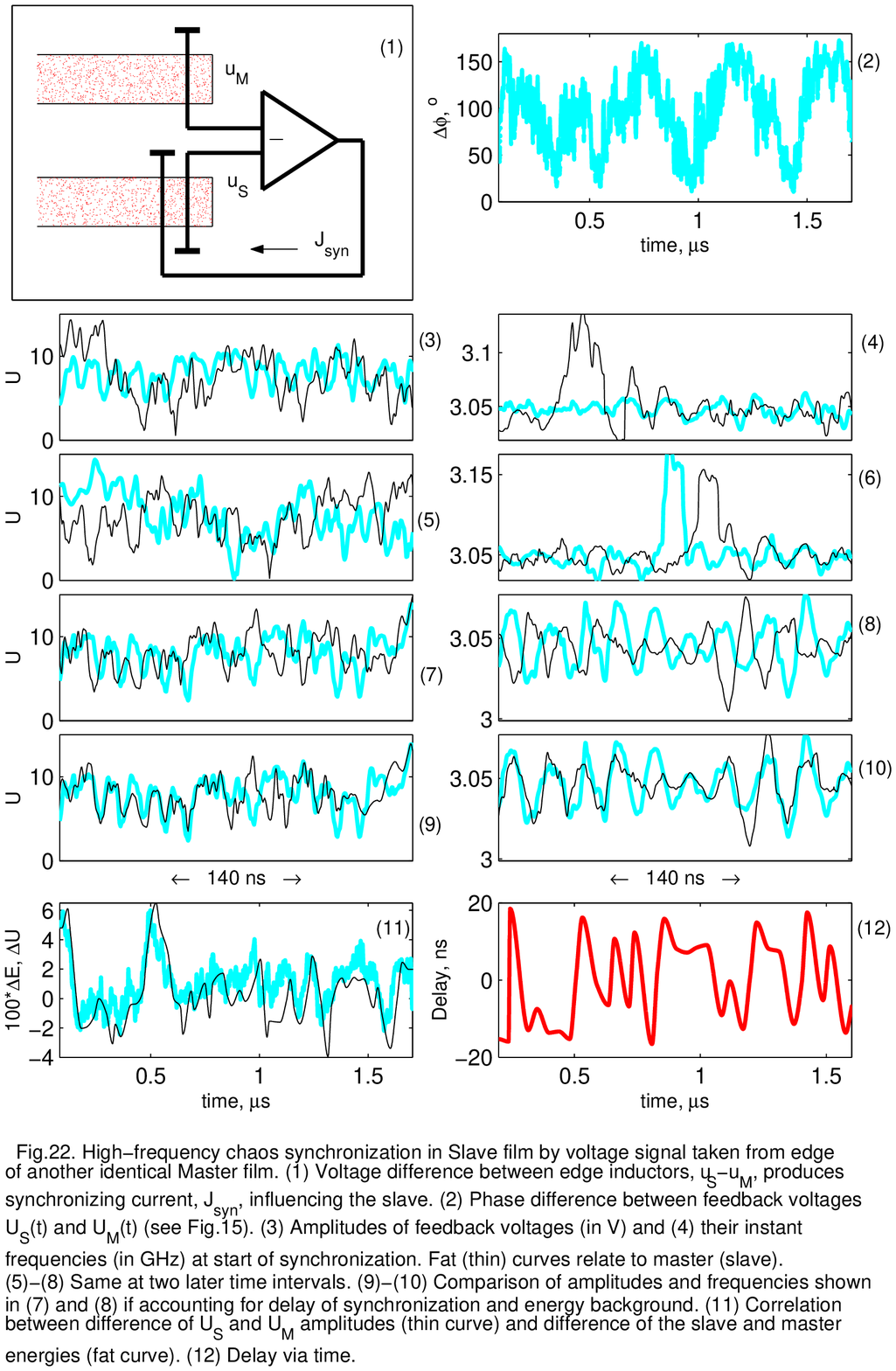}
\end{figure}

Paradoxically, because of almost ideal coherence of chaotic auto-generation,
at available duration of numerical experiments, high-frequency (carrier)
synchronization is automatically executed and thus calls no problem. The
problem consists of synchronization of phase and amplitude chaotic
modulation.

Films with \,$\,\,\,w/\lambda =3\,$\, were investigated. Various
linear and
non-linear (saturated) connections between \,$\,u_{S}-u_{M}\,\,$\,\ and \,$\,J_{syn}\, \,$\,%
\ were tested, but none of variants has shown an evident superiority.
Typical results are illustrated by plots (3)-(8) in Fig.22 which show
three pieces of synchronization with respect to envelopes of feedback
voltages in the master and slave, \,$\,U_{M}(t)\,\,$\,\ and
\,$\,U_{S}(t)\,\,$\,\ (see scheme in Fig.15), and to their instant
frequencies.

Undoubted signs of partial synchronization can be noticed already in 100 ns
after beginning of the process. They become better visible if take into
account obvious delay of the slave response. Plots (9) and (10) show how the
picture looks if correction by delay is made. Most of details of signal \,$\,%
U_{M}(t)\,\,$\,\ circulating in feedback device of master generator
are reproduced by similar signal \,$\,U_{S}(t)\,\,$\,\ in the slave.
According to plot (2), high-frequency (carrier) phase difference
between these signals does not leave the frame \,$\,\,\pm
80^{o}\,\,$\,. Nevertheless, in respect to the modulation there are
essential quantitative and sometimes qualitative distortions. The
results are far from copying of chaos. To estimate them we need in a
more soft criterion for partial synchronization.

7.14. DEFECTS OF SYNCHRONIZATION.

Careful analysis of data reflected in Fig.22 yields that most part of
misalignment between \,$\,U_{M}(t)\,\,$\,\ and \,$\,U_{S}(t)\,\,$\,\
can be related to two reasons: (i) not constant but slowly
time-varying delay, \,$\,\Delta T(t)\,$\, , and (ii) slowly
time-varying amplitude shift which is in close correlation with
difference between energies of master and slave film, \,$\,\Delta E(t)=\,$\, \,$\,%
E_{S}(t)\,$\, \,$\,-E_{M}(t)\,$\, . The latter statement is confirmed
by plot (11). Notice that the delay, as shown at plot (12), may be
both positive and negative (which corresponds to anticipating
synchronization). This findings mean that, approximately, the
relation

\begin{equation}
U_{S}(t)\approx \xi \lbrack E_{S}(t)-E_{M}(t)]+U_{M}(t-\Delta T(t))
\end{equation}
takes place with some coefficient \,$\,\xi \,$\, . Characteristic
time scale of variations in \,$\,\Delta E(t)\,$\, , in amplitude
shift \,$\,\xi \Delta E(t)\,$\, and in the delay \,$\,\Delta
T(t)\,$\, too, are about ten times slower then variations in phase
and amplitude of\thinspace\ \,$\,U_{M,S}(t)\,$\, during a cycle of
soliton production.

Hence, the conclusion arises that some low-frequency component of the
chaotic dynamics is not sensitive to synchronizing signal and
therefore hinders to synchronize fast components which themselves are
rather sensitive. Additional surprising evidence for this is that
sharp switching from one master signal, \,$\,u_{M}(t)\,\,$\,, to
another (may be the same but shifted by a time) temporarily improves
synchronization which then becomes spoiled at longer time scale.

7.15. SYNCHRONIZATION BY FEEDBACK SIGNAL.

Seemingly, better synchronization will be achieved if synchronizing
signal is taken from the ``heart'' of generator, i.e. from the
feedback device. In the simplest variant shown by plot (A) in Fig.23,
the master feedback voltage mix up with the slave one, with some
coefficient \,$\,\,\alpha \,$\,.

In this experiment besides the control voltages, \,$\,u_{S}(t)\,\,$\, and \,$\,%
u_{M}(t)\,\,\,$\,, were under comparison taken from wire inductors
placed between active and passive loops (see plot (B)). At
\,$\,\,\alpha =0.4\,\,$\,, best
results of synchronization with respect to the power absorption, \ \,$\,%
P_{M,S}(t)\,$\, , amplitude of control voltage, \
\,$\,u_{M,S}(t)\,$\, , and amplitude and instant frequency of
feedback voltage, \ \,$\,U_{M,S}(t)\,$\, , are presented by four
center plots in Fig.23. Again, here slow time-varying delay between
Slave and Master signals and slowly difference in their magnitudes
were found, in correspondence with Eq.5. Similar correlation between
difference in power absorption and difference in film energies is
detected in plot (C).

Plot (F) allows to see what the events develop in a concrete point
marked by star at plot (B) (which itself represents a snapshot of
quasi-local energy pattern). The new interesting observation follows
from the histogram (probability distribution) of delay, at plot (E).
Sharp peaks in this plots are separated by \,$\,\,\approx
0.35\,\,$\,\ ns which is close to period of precession. This may be
hint on fine features of synchronization still to be understood.

Clearly, a kind of rough synchronization takes place, but in details it
looks even worse than in the above circumferential case.

Therefore stronger mixing with \,$\,\,\alpha =0.8\,\,$\,\ was tested.
Its results are shown in Figs.24a-b (see captures to these figures).
According to top four plots in Fig.24a, results occurs definitely
better than at \,$\,\,\alpha =0.4\,\,$\,. This improvement is
directly connected to essentially smaller difference between carrier
phases of \ \,$\,U_{S}(t)\,$\, and \ \,$\,U_{M}(t)\,$\, (plot (4) in
Fig.24b) than in previous case (plot (D) in Fig.23).

\begin{figure}
\includegraphics{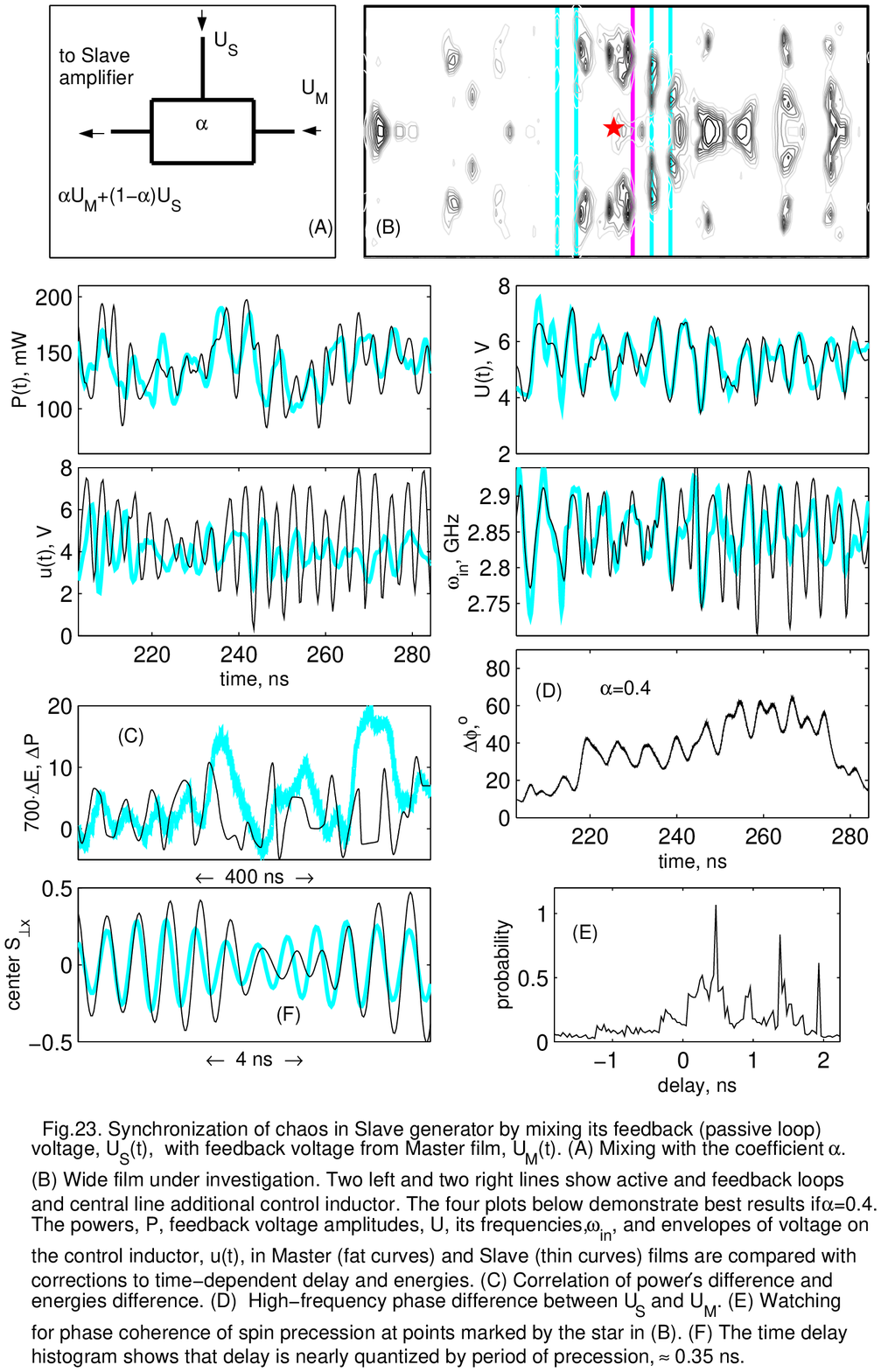}
\end{figure}

\begin{verbatim}
\end{verbatim}

7.16. MORE ABOUT\ DEFECTS OF SYNCHRONIZATION.

From the power and feedback voltage time series it is seen that most of
maxima and minima corresponding to soliton births are well reproduced by
slave system. However, synchronization of events in the interspace between
driving and feedback inductors is rather bad as registered by the powers, \ \,$\,%
P_{M,S}(t)\,$\, , and control voltages, \,$\,u_{M,S}(t)\,$\, , in
Fig.24a . This means that magnetization pattern in slave film
imitates that in master film but never copies it.

This misalignment can be explained as follows. There is no rigid
connection between film energy and a number of solitons. Some
background (smoothly distributed and slowly varying) part of energy
behaves more or less autonomously and remains out of synchronization.
Plots (1)-(5) in Fig.24b demonstrate that all the characteristics of
instant (time-local) quality of synchronization are closely
correlated with difference between the energy backgrounds in slave
and muster films. Plot (C) shows that (in concrete case under
consideration) characteristic frequency of chaotic variations of
energy background is about 30 MHz, which is of order of inverse
relaxation time, \,$\,\,\Gamma ^{-1}\,\,$\,.

Fig.24c presents one more example of the same kind of synchronization, with \,$\,%
\,\alpha =0.9\,\,$\,\ , at wide film (\,$\,w/\lambda =5\,$\,) in
strongly chaotic regime. Under corrections to influence by
non-synchronized energy background, the picture of synchronization
with respect to power absorption looks enough good (plot (D)). But
the whole picture has no significant differences from previous case.

7.17. MAGNETIC\ TURBULENCE.

The Figs.25a-b illustrate what happens at too powerful
auto-generation which is achieved under weak output saturation,
\,$\,I_{sat}=100\,\,$\,\ mA, although at
relatively small linear gain (=4). In spite of that film is narrow (\,$\,%
w/\lambda =2.5\,$\,), extremely chaotic shapeless energy pattern is
formed by irregularly scattered solitons (see top picture in
Fig.25b).

In this case (which may be named turbulent magnetic chaos), noticeable
chaotic drift (diffusion) of the carrier phase takes place due to phase
slips, as shown at plot (A) in Fig.25a. Plots (F)-(H) and (C)-(E) present
well expressed slips in local magnetization and in such the integral
characteristics as feedback voltage. According to plot (B), corresponding
slow chaotic variations of carrier frequency are excellently anti-correlated
with the energy (as the non-isochronity implies).

\begin{figure}
\includegraphics{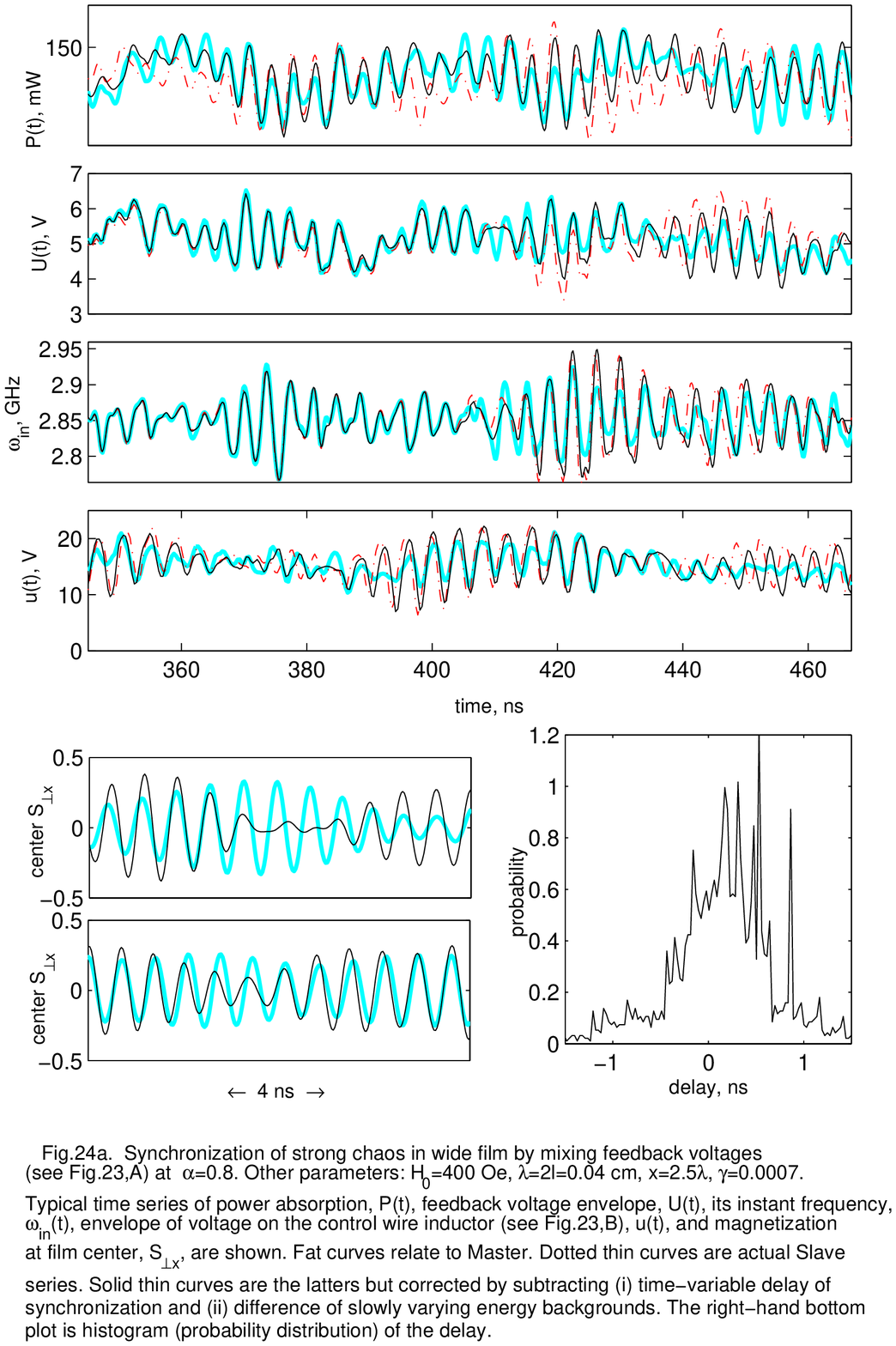}
\end{figure}
\begin{figure}
\includegraphics{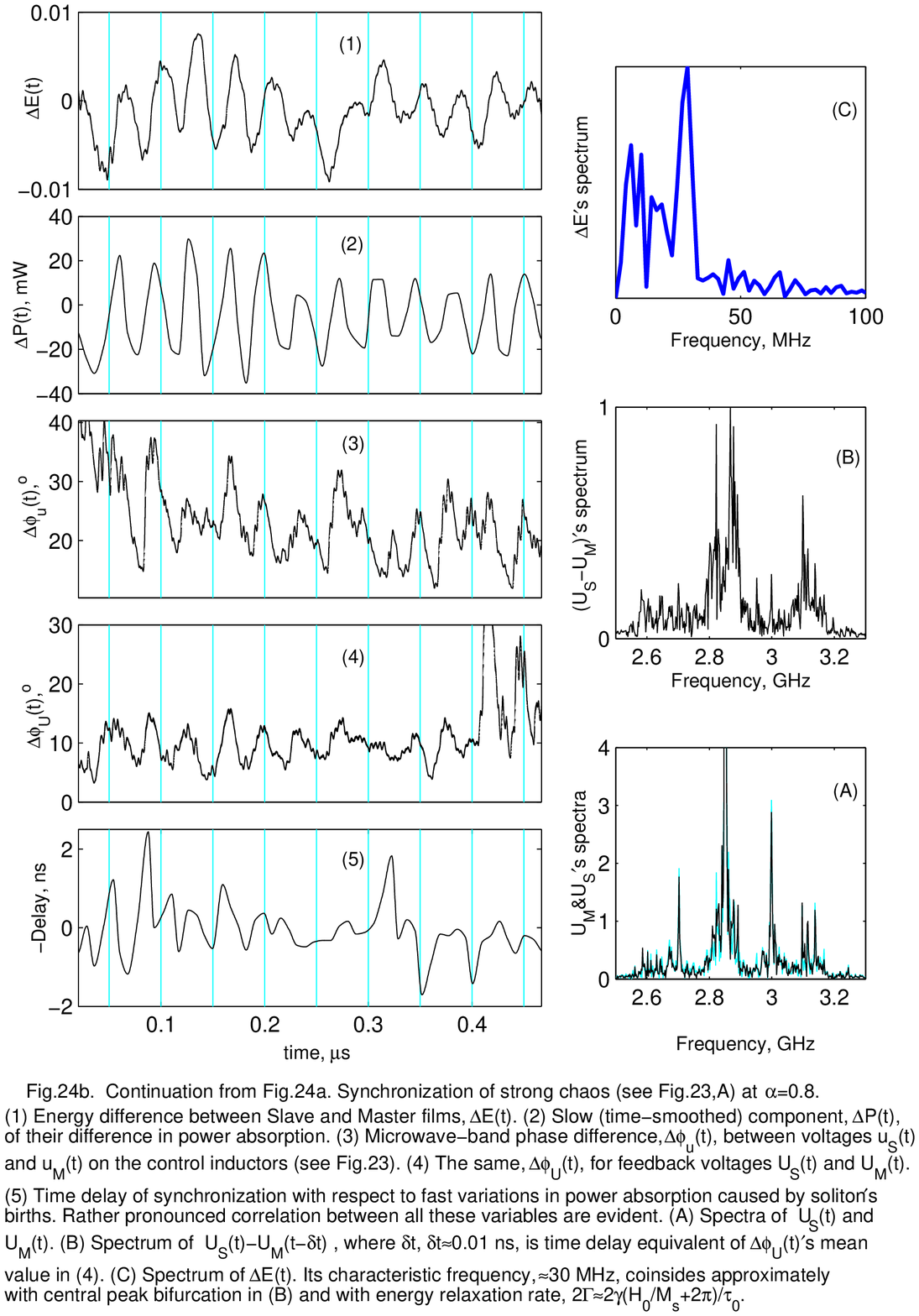}
\end{figure}
\begin{figure}
\includegraphics{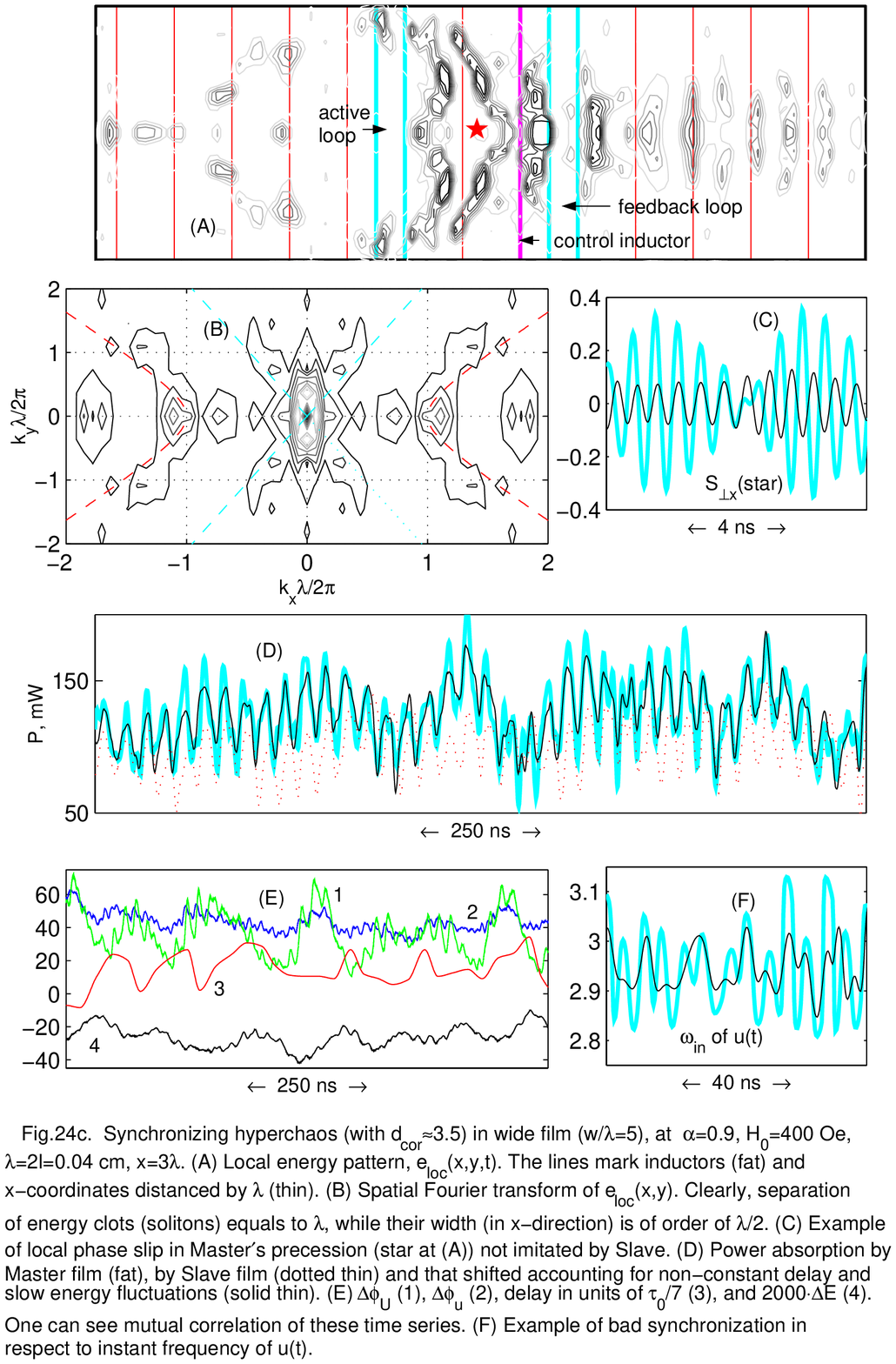}
\end{figure}

Two interest things should be appointed.

(i) In contrast to commonly strong chaos, with \,$\,\,d_{cor}\approx 3.9\,\ \,\,$\,%
if estimated from feedback voltage time series (see plot (C) in
Fig.25b), the energy behaves much more regularly and can be
characterized by own fractal dimension, \,$\,\,d_{cor}\approx
2.7\,\,$\,. Possibly, this is just manifestation of above mentioned
autonomy of the energy background.

(ii) Very slow chaotic variations in energy (and thus in the carrier
frequency and other related values) are observed, with characteristic
frequencies down to 3 MHz (plots (D) and (E) in Fig.25b).

7.18. CHAOTIC PULSE AUTO-GENERATION.

Slight decrease in linear gain (from 4 to 3) at the same film at same
output saturation results in new interesting regime of
auto-generation which is characterized by sharp pulses of the power
absorption,\thinspace \,$\,\,\,P(t)\,$\, , and voltages in feedback
and driving loops, \,$\,\,U(t)\,\,$\,\ and \,$\,\,V(t)\,$\,.

This example is illustrated by Figs.26a-b. Theoretically, \,$\,P=\,$\, \,$\,%
\,\left\langle JV-IU\right\rangle _{T}\,\,$\, , where
\,$\,I(t)\,\,$\,\ and \,$\,J(t)\,\,$\,\
are currents in feedback and driving loops (see scheme in Fig.15), and \,$\,%
\left\langle ..\right\rangle _{T}\,$\, means averaging over period of
precession. The part \,$\,\,\left\langle JV\right\rangle _{T}\,$\, \
is power absorption from driving loop which can be written as
\,$\,\,\left\langle JV\right\rangle _{T}=\,$\, \,$\,|J||V|\cos (\phi
_{V}-\phi _{J})\,\,$\,\ , with \,$\,|J|\,$\, and \,$\,|V|\,\,$\,\
being envelopes and \,$\,\phi _{V}\,\,$\,\ and \,$\,\,\phi
_{J}\,\,$\,\ phases. Fat curve and thin curve close to it at plot (5)
in Fig.26b show pulsations of \thinspace \,$\,\,P\,$\, and
\,$\,\,\left\langle JV\right\rangle _{T}\,\,$\,\ , respectively,
while the low-amplitude curve here represents \,$\,\,\left\langle
IU\right\rangle _{T}\,\,$\,\ and shows that the feedback loop is
consuming film's energy.

In this case, again (i) essential difference between common chaos (with \,$\,%
\,d_{cor}\approx 3.3\,$\, )\,$\,\,\,$\, and particular energy chaos (with \,$\,%
\,d_{cor}\approx 2.2\,\,$\,), and (ii) slow energy variations (with
characteristic frequencies \,$\,\,\sim 2\ \,$\,\ MHz if not less)
\,$\,\,\,$\,take place.

At last, due to relatively simple and expressively shaped chaos and
relatively small fractal dimension in this case, we may try do draw
three-dimensional projection of corresponding chaotic attractor. Two
variants of the projection are shown in Fig.28.

7.19. ON\ POWER\ THRESHOLD OF\ CHAOS.

In numerical simulations, we can set any gain and output power as we want.
In practice, small-sized high-frequency amplifiers have rather limited
output power. Therefore, consider rough analytical estimate of the smallest
(threshold) power absorption by film which is sufficient for chaotic
auto-generation.

\begin{figure}
\includegraphics{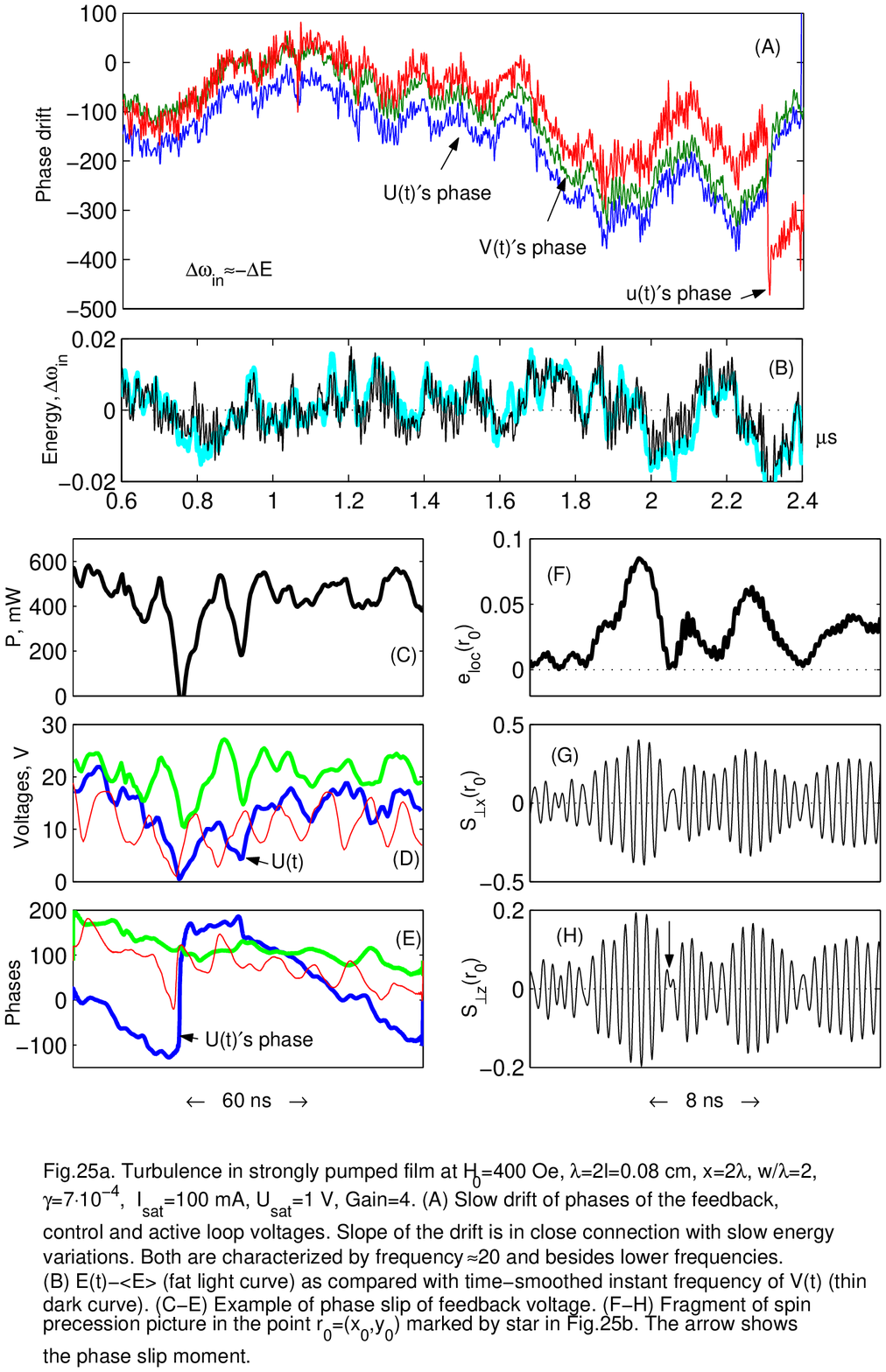}
\end{figure}
\begin{figure}
\includegraphics{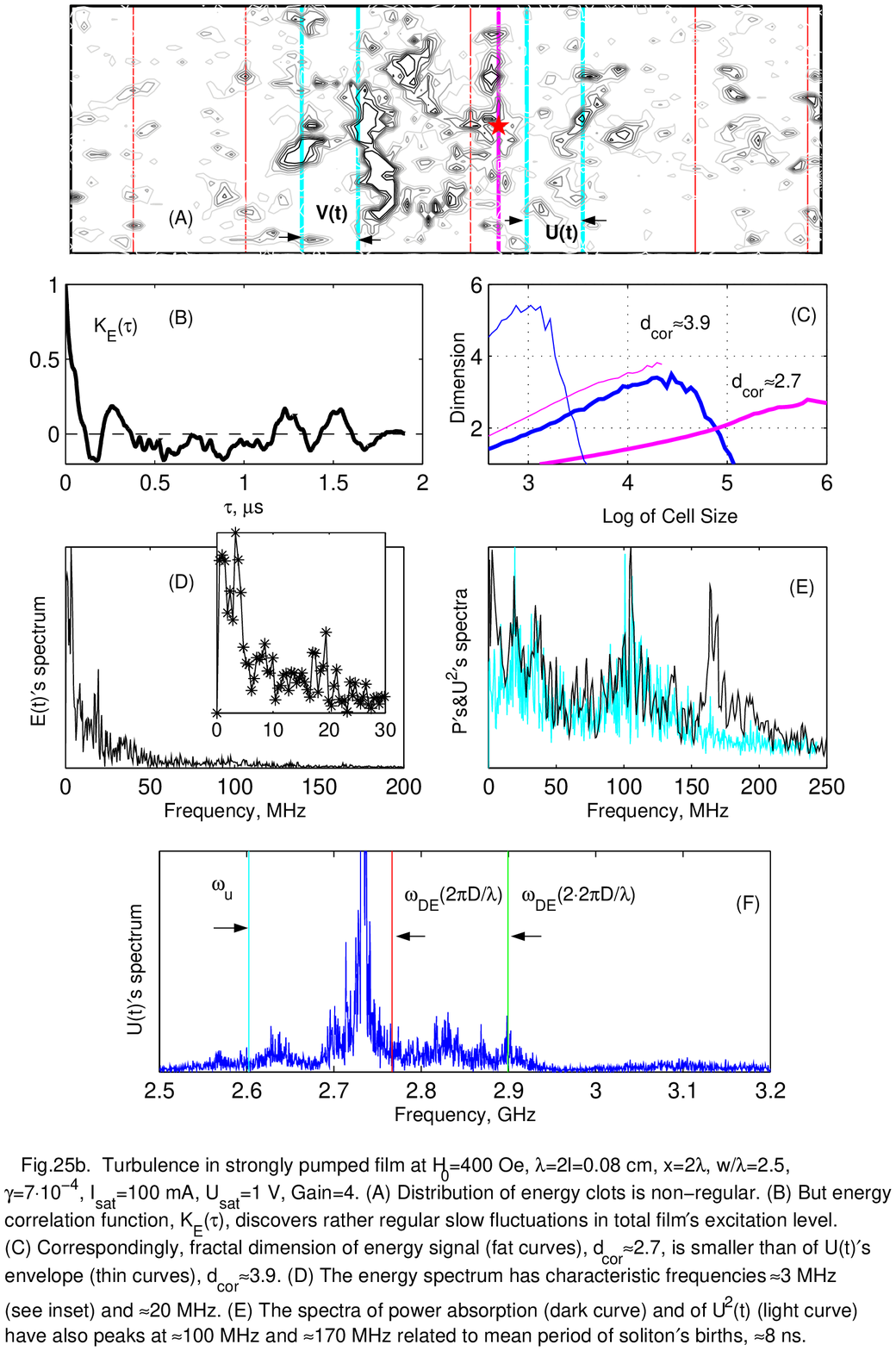}
\end{figure}

Time averaging of the energy conservation law (6.3), if combined with
the relations (6.4) (at \,$\,E_{0}=0\,$\, as it was seen) and (6.6)
(i.e. for not too short-wave chaos), yields

\begin{equation}
\left\langle P\right\rangle _{T}\approx H_{0}\Gamma \left\| S_{x}\right\|
^{2}\approx H_{0}p\Gamma |\Psi |^{2}=\omega _{u}\Gamma |\Psi |^{2}\,\ ,
\end{equation}
where \,$\,\ \Psi \,\,$\,is the wave function considered in Sec.5,
\,$\,\,p\,\,$\,\ is eccentricity of uniform precession,
and\thinspace\ \,$\,\left\langle P\right\rangle _{T}\,$\, \thinspace
is mean power absorption per unit volume (in
dimensionless units). Hence, in according with the estimates of the minimum \,$\,%
|\Psi |\,$\,'s \ value, \,$\,\,A_{\min }\,\,$\,, which initiates wave
instability (see Sec.5.5), we can estimate power level (per unit
volume) which is sufficient for beginning of chaos as

\begin{equation}
\min \,\left\langle P\right\rangle _{T}\approx \omega _{u}\Gamma
^{2}/|\varkappa |
\end{equation}
In real units, for the total threshold power absorption,
\,$\,P_{thr}\,\,$\,\ , in a film with volume \,$\,V\,\,$\,\ this
formula implies

\begin{equation}
P_{thr}\approx 2\pi M_{s}^{2}Vf_{u}\Gamma ^{2}/|\varkappa |\,\ ,\,
\end{equation}
where \,$\,\,f_{u}\,\ \,$\,is uniform precession frequency (in \,$\,\,\,$\,s\,$\,^{-1}\,$\,) and \,$\,%
\,\Gamma \,\,$\,remains dimensionless.

In YIG film, at \,$\,\,H_{0}\approx 550\,\,\,\,$\,Oe\
(i.e.\,$\,\,\approx 4\,\,$\,\ in
dimensionless form) \,$\,\,|\varkappa |\approx 1.4\,\,$\,\ and\thinspace\ \,$\,%
\,f_{u}\approx 3.2\cdot 10^{9}\,\,\,$\,s\,$\,^{=1}\,$\,. If take \,$\,\,\Gamma =\,$\, \,$\,0.007\,\,$\,%
\thinspace , which corresponds to approximately 1 Oe half-width of linear
FMR, and the volume be \,$\,\,7\,\,$\,mm\,$\,\,\,$\,\ \,$\,\times 2\,\,$\,mm \,$\,\times 10\,\,$\,%
micron\thinspace\ then Eq.8 gives \,$\,\,P_{thr}\approx 180\,\,$\,\
mW. However, the estimate is rather sensitive to energy losses, and
at 0.5 Oe half-width of FMR it turns into \,$\,\approx 40\,\,$\,\ mW.
Probably, this is still overestimated value since not the whole
volume must be critically excited at threshold pump.

%!8

%\vspace{2 cm}

\,\,\,

\section*{Conclusion}

We have reported selected results of work %
which was made in 2002-2003 years. %
Additional information about this work, - in the form %
of HTML presentation (also made at that time) %
with figures, movies and comments %
to them, - can be obtained by downloading and unzipping %
file\,\,  %
\url{http://yuk-137.narod.ru/experience/mfs_2002.zip} %
\,. This presentation includes also some information %
about our (together with our colleagues) first %
attempts of practical generation and synchronization of %
magnetic-wave chaos and its properties in comparison with that %
observed in numerical simulations. %

On the latter subjects and, besides, %
physical properties of our %
ferrite (YIG) films and %
first practical applications of electro-magnetic chaos, %
generated with the help of these films, %
to direct microwave-frequency secure communication, %
see also our recent journal articles [1,2].

\,\,\,

Principal conclusions what can be extracted from the aforesaid %
material are as follow. %

(i) There are two main machanisms of the magnetic-wave chaos %
auto-generation. One is creation of non-regular chains of (``gray'') %
envelope solitons composed (in in-plane magnetized films under our %
above consideration) primarily with `surface'' (``Damon-Eshbach'') %
magnetostatic spin waves (MSW). %
Another mechanism is parametric energy transfer from these waves %
to relatively short ``backward'' MSW (``bulk'' MSW) and reversed %
process.

(ii) The first mechanism dominates at not too large (in-plane) %
magnetizing field (therefore,  generates lower carrier frequencies) %
but it needs in much smaller level of energy consumption %
(dissipation) in film and, hence, much smaller amplification in a feedback %
circuit. Besides, importantly, it produces much more %
``rich'' chaos, which allows, in principle, to ``hide'' %
a large amount of information.

(iii) However, these advantages of the ``parametric'' chaos %
(as compared with the ``solitonic'' one) are accompanied %
with probable instability of the parametric interaction between %
surface and backward MSW, which may take form of explosive %
and irreversible transfer of energy from surface to short %
backward waves. A degree of the irreversibility is as large as %
great is number of excited backward wave modes per one exciting %
surface mode. At some regimes of auto-generation one can observe %
from time to time fast and practically full energy swaps into short
backward waves and, consequently, sharp and long-standing %
suppressions of auto-generation (sometimes, under definite %
conditions, it may break for periods up to nearly \,$\,\mu $s\,). %

This phenomenon (in the presentation, we called it %
``parametric collapse'') was observed both in numerical %
and real experiments (with good correspondence between %
them, though numeric procedures inevitably loss most short waves). %
Naturally, it can create strong obstacles to %
synchronization of chaos, because the breaks of %
auto-generation imply uncontrollable and hardly %
reproducible in ``slave'' system) phase slips. %

(iv) Therefore, to avoid the ``parametric collapse'' but, %
at the same time, exploit parametric processes as much %
as possible, it is necessary to introduce some auto-regulation %
of feedback amplification. It can help in %
generation of such output electromagnetic signals what %
possess not too strong amplitude modulation, at %
strong enough  phase and frequency chaotic modulation, %
and cause satisfactory ``synchronous chaotic response'' %
in slave system.

(v) Fractal (correlation) dimensions of both high-frequency %
chaotic oscillations of different variables (local %
magnetization values, e.m.f.'s in antennae, %
and voltages and currents in feedback circuit) and %
low-frequency variables (amplitudes, phases and ``instant %
frequencies'') almost always lie in interval from 2 to 4 %
and usually between 3 and 4, thus demonstrating ``hyper-chaos''. %
This numeric result well agrees with fractal dimensions of %
low-frequency (amplitude and phase) signals obtained from real %
chaotic generators. At the same time, some of low-frequency %
variables, both real and numerical (e.g `quasi-local energy %
density'' and total ``excess'' magnetic energy) can show %
dimensions nearly by unit lesser, from interval 2$\div $3\,. %

(vi) These facts, seemingly,  help to understand %
why, in spite of not high fractal dimension of the our chaos, %
we never observe ``ideal synchronization'', instead %
seeing only what can be termed ``generalized synchronization''. %
The matter is that, in the schemes under investigation, %
all the control and feedback signal are directly coupled %
with the ``surface'' (``Damon-Eshbach'') MSW modes only. %
This is obvious from comparison of voltage-current spectra, - %
whose frequency band always lies near or higher the uniform %
precession frequency, - and numerically found spectra %
of local magnetization values (local precession) %
which can be concentrated mainly at twice lower %
frequencies or even four times lower ones (because of two %
successive parametric divisions of frequency), down to absolute %
lower bound of MSW spectrum (moreover, sometimes, - e.g. after  %
the parametric collapse, - become concentrated  %
just near this bound).

Therefore, it is not surprising %
if one of 4 effective relevant degrees of freedom %
of our system (the number 4 follows from fractal %
dimension between 3 and 4), - which is responsible for  %
magnitude of short backward MSW and their energy exchange %
with surface MSW, - stays far out of control we used. %

If this is true explanation of the generalized  %
character of synchronization, then we have chances %
to improve the described results.

(vii) It appeared undoubt that even rather rough %
numeric simulations are useful not only for adequate theoretical %
of real experiments, but also for their planning.

\,\,\,

Clearly, the described work leaved many %
unanswered questions and hypotheses waiting for %
numeric and experimental examination. %
This will subject of separate papers.

\,\,\,

\,\,\,

{\it REFERENCES}

\,\,

[1]\, N.I.\,Mezin,\, Pis'ma v ZhTF\, {\bf 37}, No.23, 61 %
(2011) (in Russian; translated to English by AIP).

[2]\, N.I.\,Mezin, A.A.\,Grishchenko, and Yu.E.\,Kuzovlev,\, %
Pis'ma v ZhTF\, (in press).

%%%%%%%%%%%%

\newpage

\topmargin -2 cm

\begin{figure}
\includegraphics{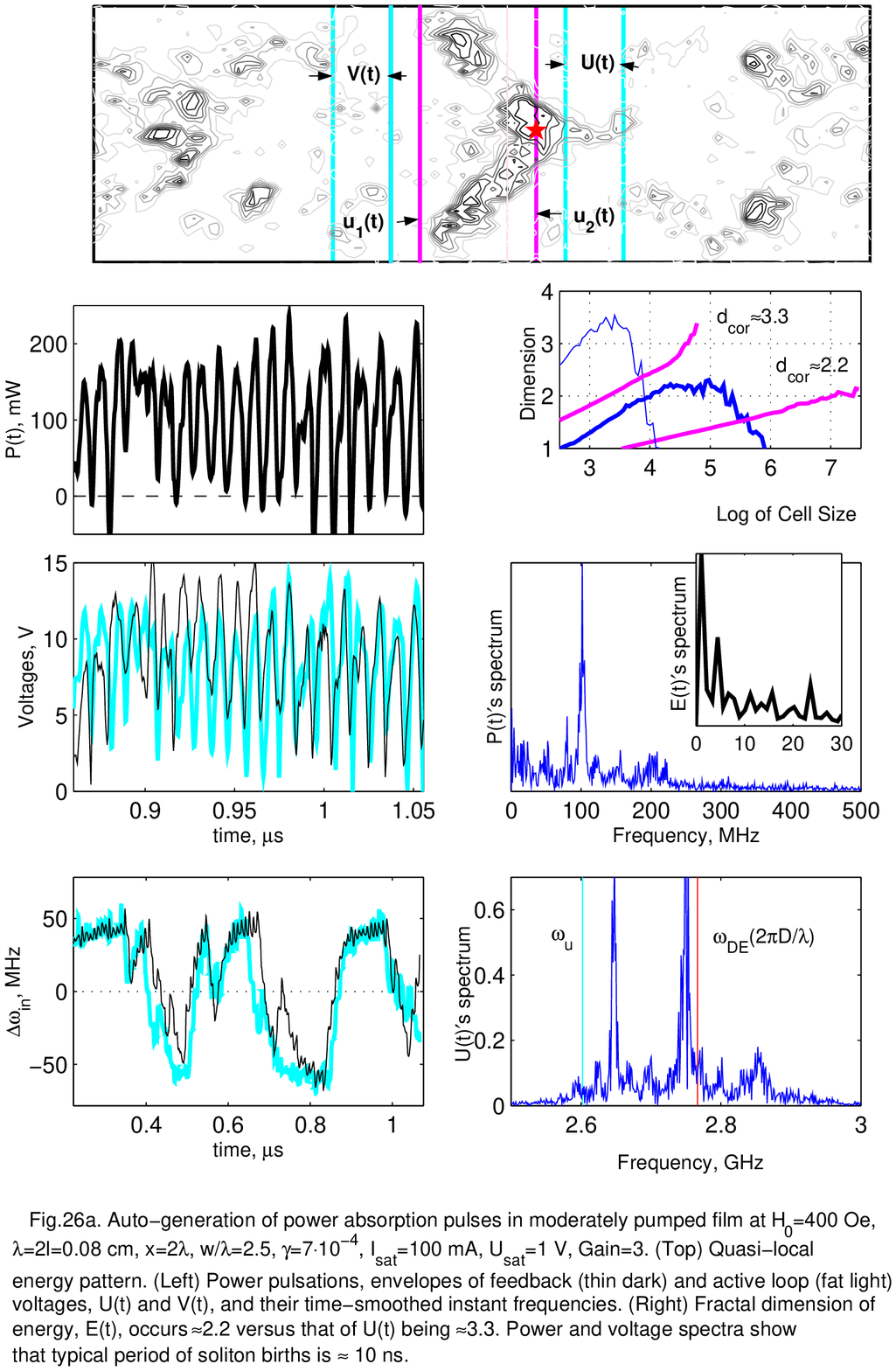}
\end{figure}
\begin{figure}
\includegraphics{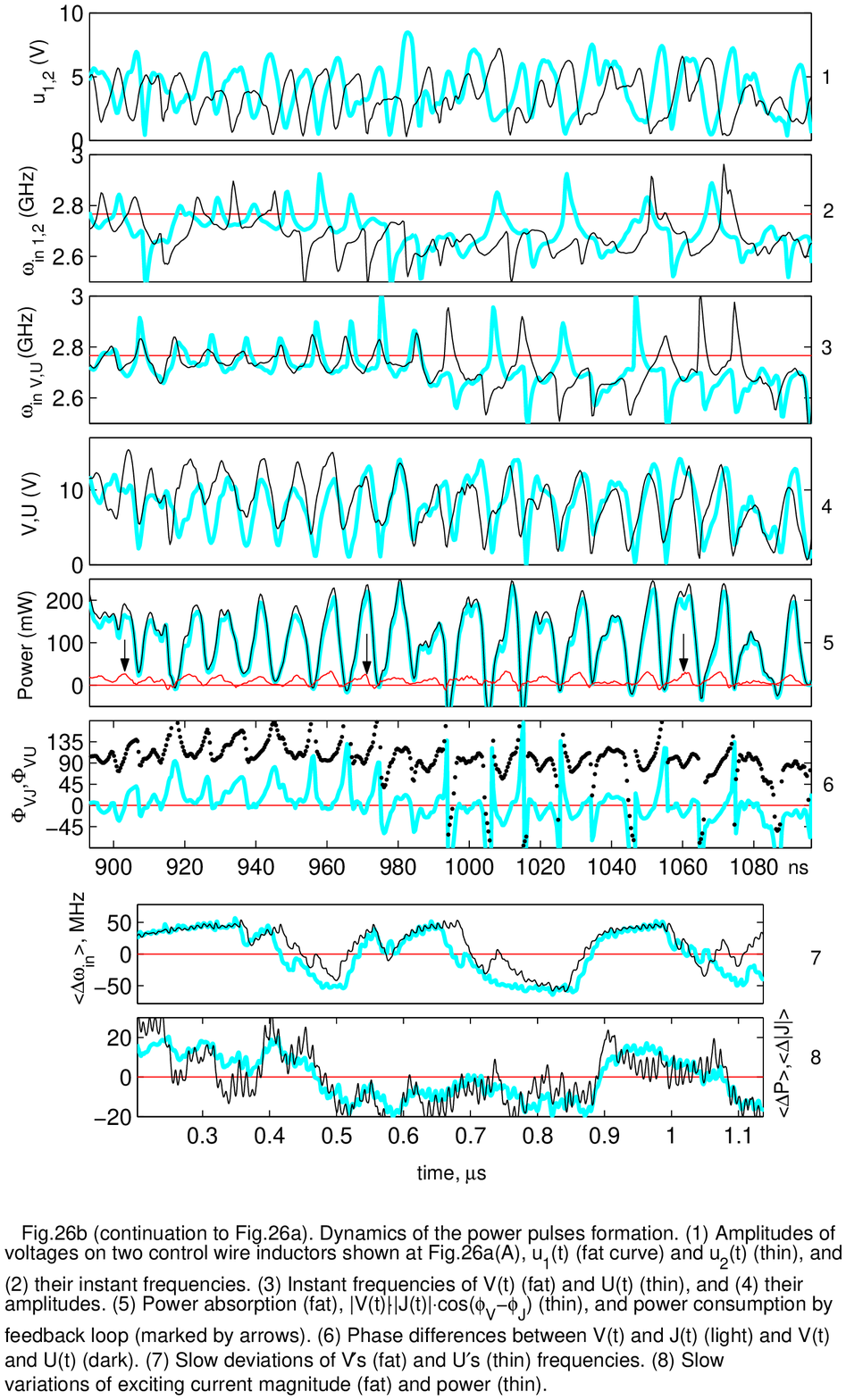}
\end{figure}

\begin{figure}
\includegraphics{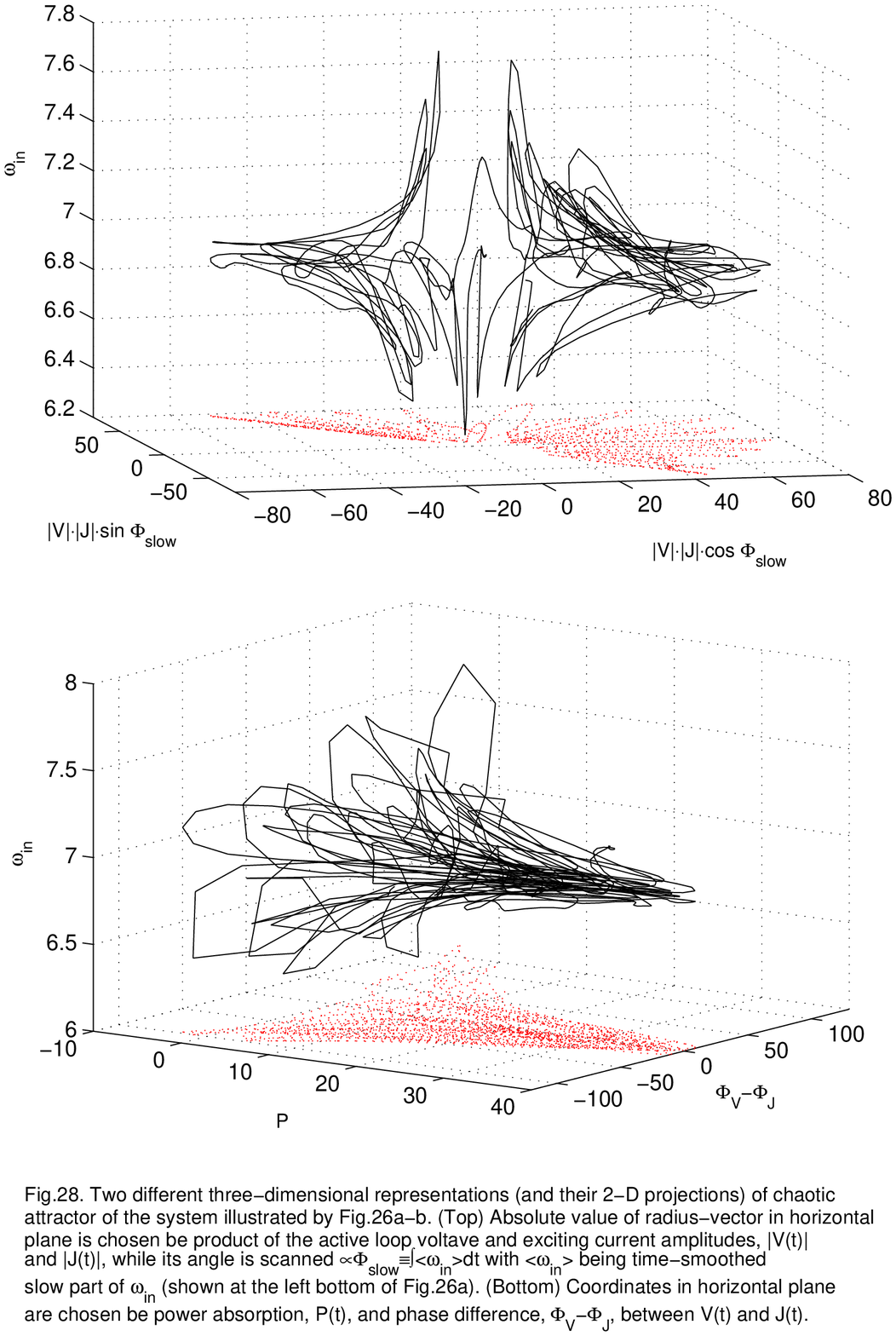}
\end{figure}

%%%%%%%%%%%

%%%%%%%%%%%%

\end{document}